\documentclass[aps,prl,twocolumn,nopacs,showpacs,superscriptaddress]{revtex4}

\usepackage{graphicx}  
\usepackage{dcolumn}   
\usepackage{bm}        
\usepackage{amssymb}   
\usepackage{amsmath}
\usepackage{units}
\usepackage[ansinew]{inputenc}
\usepackage[dvipsnames]{xcolor}

\newcommand{\ket}[1]{\left\vert#1\right\rangle}
\newcommand{\bra}[1]{\left\langle#1\right\vert}

\newcommand{\sandwich}[3]{\left< #1 \vphantom{#2 #3} \right| #2 \left|\vphantom{#1 #2} #3 \right>}

\usepackage[type1]{libertine}
\usepackage{textcomp}
\usepackage[scaled=.85]{beramono}
\usepackage[libertine,cmintegrals,cmbraces,vvarbb,slantedGreek]{newtxmath}
\usepackage[scr=boondoxo]{mathalfa}
\usepackage{bm}
\usepackage[lf]{carlito}

\usepackage{braket}
\usepackage{bm}

\makeatletter
\DeclareRobustCommand{\cev}[1]{%
  \mathpalette\do@cev{#1}%
}
\newcommand{\do@cev}[2]{%
  \fix@cev{#1}{+}%
  \reflectbox{$\m@th#1\vec{\reflectbox{$\fix@cev{#1}{-}\m@th#1#2\fix@cev{#1}{+}$}}$}%
  \fix@cev{#1}{-}%
}
\newcommand{\fix@cev}[2]{%
  \ifx#1\displaystyle
    \mkern#23mu
  \else
    \ifx#1\textstyle
      \mkern#23mu
    \else
      \ifx#1\scriptstyle
        \mkern#22mu
      \else
        \mkern#22mu
      \fi
    \fi
  \fi
}
\makeatother

\begin{document}

\title{Tunneling dynamics between superconducting bound states at the atomic limit}

\author{Haonan Huang}
\affiliation{Max-Planck-Institut f\"ur Festk\"orperforschung, Heisenbergstraße 1,
70569 Stuttgart, Germany}
\author{Ciprian Padurariu}
\affiliation{Institut für Komplexe Quantensysteme and IQST, Universität Ulm, Albert-Einstein-Allee 11, 89069 Ulm, Germany}
\author{Jacob Senkpiel}
\affiliation{Max-Planck-Institut f\"ur Festk\"orperforschung, Heisenbergstraße 1,
70569 Stuttgart, Germany}
\author{Robert Drost}
\affiliation{Max-Planck-Institut f\"ur Festk\"orperforschung, Heisenbergstraße 1,
70569 Stuttgart, Germany}
\author{Alfredo Levy Yeyati}
\affiliation{Departamento de F\'{\i}sica Te\'orica de la Materia Condensada and
Condensed Matter Physics Center (IFIMAC), Universidad Aut\'onoma de Madrid, 28049 Madrid, Spain}
\author{Juan Carlos Cuevas}
\affiliation{Departamento de F\'{\i}sica Te\'orica de la Materia Condensada and
Condensed Matter Physics Center (IFIMAC), Universidad Aut\'onoma de Madrid, 28049 Madrid, Spain}
\author{Bj\"orn Kubala}
\affiliation{Institut für Komplexe Quantensysteme and IQST, Universität Ulm, Albert-Einstein-Allee 11, 89069 Ulm, Germany}
\author{Joachim Ankerhold}
\affiliation{Institut für Komplexe Quantensysteme and IQST, Universität Ulm, Albert-Einstein-Allee 11, 89069 Ulm, Germany}
\author{Klaus Kern}
\affiliation{Max-Planck-Institut f\"ur Festk\"orperforschung, Heisenbergstraße 1,
70569 Stuttgart, Germany}
\affiliation{Institut de Physique, Ecole Polytechnique Fédérale de Lausanne, 1015 Lausanne, Switzerland}
\author{Christian R. Ast}
\email[Corresponding author; electronic address:\ ]{c.ast@fkf.mpg.de}
\affiliation{Max-Planck-Institut f\"ur Festk\"orperforschung, Heisenbergstraße 1,
70569 Stuttgart, Germany}

\date{\today}

\begin{abstract}
  Despite plenty of room at the bottom, there is a limit to the miniaturization of every process. For charge transport this is realized by the coupling of single discrete energy levels at the atomic scale. Here, we demonstrate sequential tunneling between parity protected Yu-Shiba-Rusinov (YSR) states bound to magnetic impurities located on the superconducting tip and sample of a scanning tunneling microscope at 10\,mK. We reduce the relaxation of the excited YSR state to the bare minimum and find an enhanced lifetime for single quasiparticle levels. Our work offers a way to characterize and to manipulate coupled superconducting bound states, such as Andreev levels, YSR states, or Majorana bound states.
\end{abstract}

\maketitle


In conventional superconductivity, Cooper pairs, formed out of pairs of electrons with opposite spin and momentum, condense to a macroscopic quantum state. Excitations from this ground state are Bogoliubov quasiparticles, coherent superpositions of an electron and a hole. In a uniform superconductor, these excitations occupy delocalized states in a continuum separated from the condensate by an energy gap $\Delta$. Localized Bogoliubov states \cite{pillet2010,bretheau2013,janvier2015}, however, can be coupled to engineer quantum systems predicted to show parity protection (even/odd particle number conservation) as well as topological protection \cite{Hassler2011,beenakker2016}, with applications in quantum information processing \cite{mourik_signature_2012,Nadj-Perge2014}.

Bogoliubov quasiparticles can be localized at the atomic scale in Yu-Shiba-Rusinov (YSR) states created within the gap by perturbing the superconductor with a magnetic impurity \cite{Yu1965,Shiba1968,Rusinov1969}. YSR states have been observed on a variety of magnetic atoms -- intrinsically present or deliberately placed on the surface -- on various superconducting substrates using scanning tunneling microscopy (STM) \cite{Yazdani1997,Franke2011,Cornils2017,Senkpiel2018a,heinrich_single_2018}. Due to its capabilities to resolve and manipulate single atoms \cite{Heinrich2004,Meier2008,Nadj-Perge2014,Natterer2017,Senkpiel2018}, the STM is ideally suited to realize a tunable coupling between one YSR state at the tip apex and another on the sample surface. Elementary charges tunneling between these two YSR states results in subgap excitations that subsequently relax into the continuum of each superconductor acting as a reservoir. Combining tunable coupling and long YSR state lifetimes gives access to the transition from a weakly coupled sequential tunneling regime to an emergent coherent regime, where the coupling overcomes the relaxation. In this way, the tunnel current provides means to electrically characterize the states, their energy, relaxation lifetime, tunnel coupling, and their coupling to the environment \cite{VanderVaart1995,Fujisawa1998,Leggett1987}. We, thus, demonstrate the ultimate limit of miniaturization for charge transport.

\begin{figure}
\centerline{\includegraphics[width = \columnwidth]{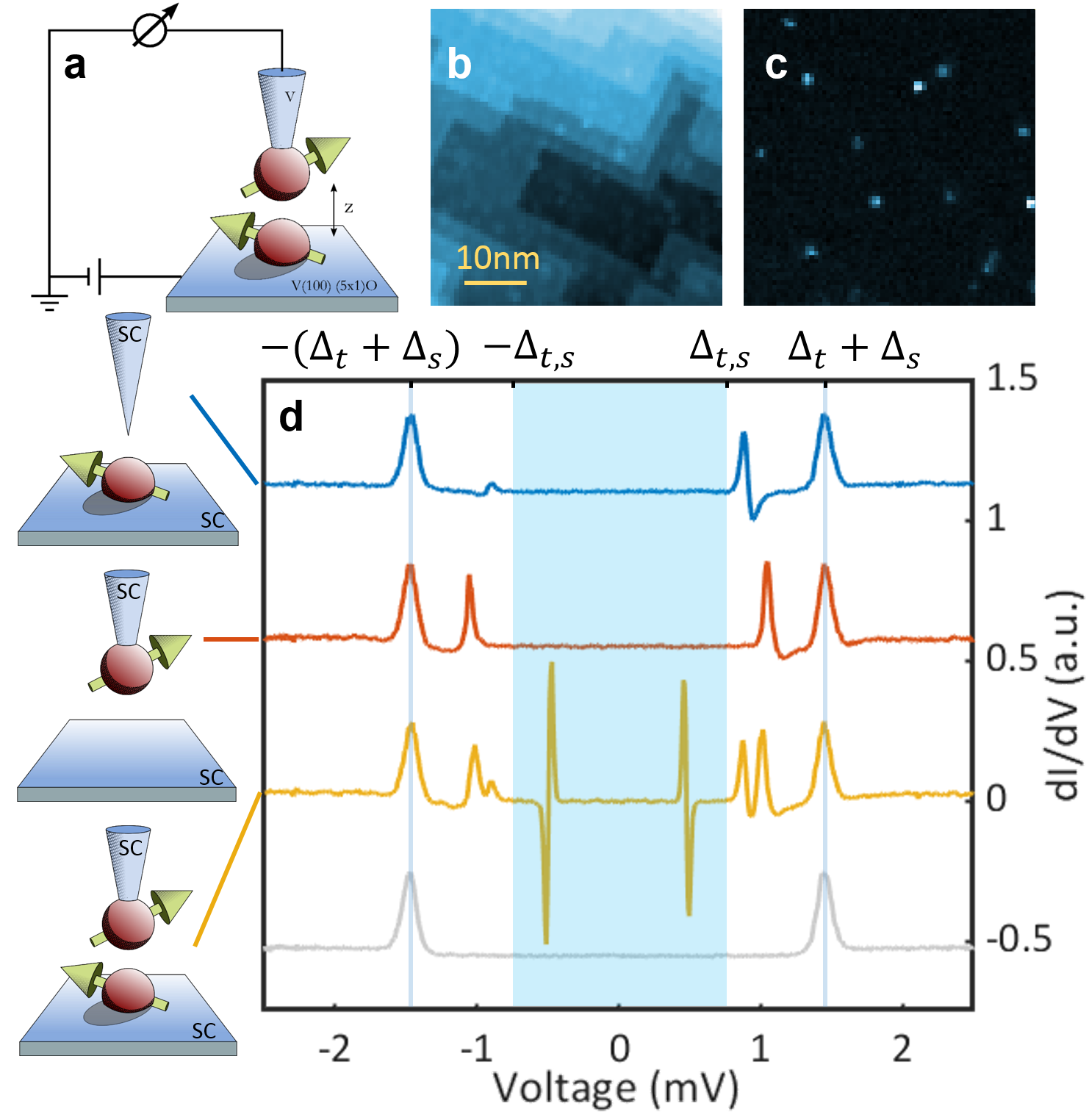}}
\caption{\textbf{Tunneling between Yu-Shiba-Rusinov states.} \textbf{a,} Schematic of the voltage-biased tunnel junction showing a YSR state at the tip apex over an intrinsic YSR state at the V(100) surface. \textbf{b,} Topographic image of the V(100) surface. \textbf{c,} Current map of the same area as in \textbf{b} at a bias voltage just below $2(\Delta_{\textrm{\small t}}+\Delta_{\textrm{\small s}})$. YSR states show up as bright spots. \textbf{d,} Differential conductance spectra at low conductance at 10\,mK. Blue: YSR state on the sample and clean (i.\ e.\ no YSR state) superconducting tip. The peaks inside the superconducting gap at $\pm(\varepsilon_{\textrm{\small s}} + \Delta_{\textrm{\small t}})$ are due to conventional YSR tunneling. Red: YSR state on the tip and clean sample. Yellow: A YSR state on the tip and another YSR state on the sample. The pair of sharp peaks inside the shaded region at $eV=\pm(\varepsilon_{\textrm{\small s}} + \varepsilon_{\textrm{\small t}})$ are Shiba-Shiba tunneling peaks. Gray: clean tip and clean sample, where coherence peaks at $eV=\pm(\Delta_{\textrm{\small t}}+\Delta_{\textrm{\small s}})$ and a clean gap in between can be seen.} \label{fig:Fig1}
\end{figure}

Here, we use an STM at a base temperature of 10\,mK \cite{Assig2013} to realize tunneling between a YSR state on the apex of a superconducting vanadium tip and a YSR state from an intrinsic impurity on the superconducting V(100) sample. A schematic of the measurement setup and a topographic map of the V(100) surface are shown in Fig.\ \ref{fig:Fig1}a and b. A sparse distribution of intrinsic magnetic impurities can be observed on the V(100) surface (see Fig.\ \ref{fig:Fig1}c), which produce a single well-defined YSR state inside the gap. The YSR states appear at a wide range of energies, indicating a varying local exchange coupling. With controlled indentation of the tip on the sample surface, we can reproducibly introduce a YSR state on the apex of the tip at defined energies inside the gap. For more details refer to the Supplementary Information \cite{supinf}.

\section*{Tunneling between Superconducting Bound States}

Typical differential conductance ($dI/dV$) spectra at a low conductance setpoint $G_\textrm{\small N}$ are shown in Fig.\ \ref{fig:Fig1}d for a YSR state in the sample (blue), in the tip (red), in both tip and sample (yellow), as well as an empty gap (grey) without any YSR state. The grey curve represents tunnelling between a clean superconducting tip and sample. We observe the two typical coherence peaks, separated by twice the sum of tip and sample gaps $2(\Delta_\textrm{\small t} + \Delta_\textrm{\small s})$. In the red and blue spectra, we observe a single pair of YSR peaks located at $\pm(\varepsilon_{\textrm{\small s,t}} + \Delta_{\textrm{\small t,s}})$, where $\varepsilon_{\textrm{\small s,t}}$ is the energy of the YSR state inside the sample gap (s) or tip gap (t) shifted by the tip or sample gap $\Delta_{\textrm{\small t,s}}=750\pm 10\,\mu$eV, respectively. This process describes tunneling from the YSR state into the continuum and vice versa, which we refer to as conventional YSR tunneling \cite{Ruby2015a}. The involvement of the gapped continuum generally results in a forbidden region for bias voltages $|eV|\le \Delta_\textrm{\small t,s}$ shown as the shaded area in Fig.\ \ref{fig:Fig1}d, where no tunneling is expected at low temperature and low conductance.

\begin{figure}
\centerline{\includegraphics[width = \columnwidth]{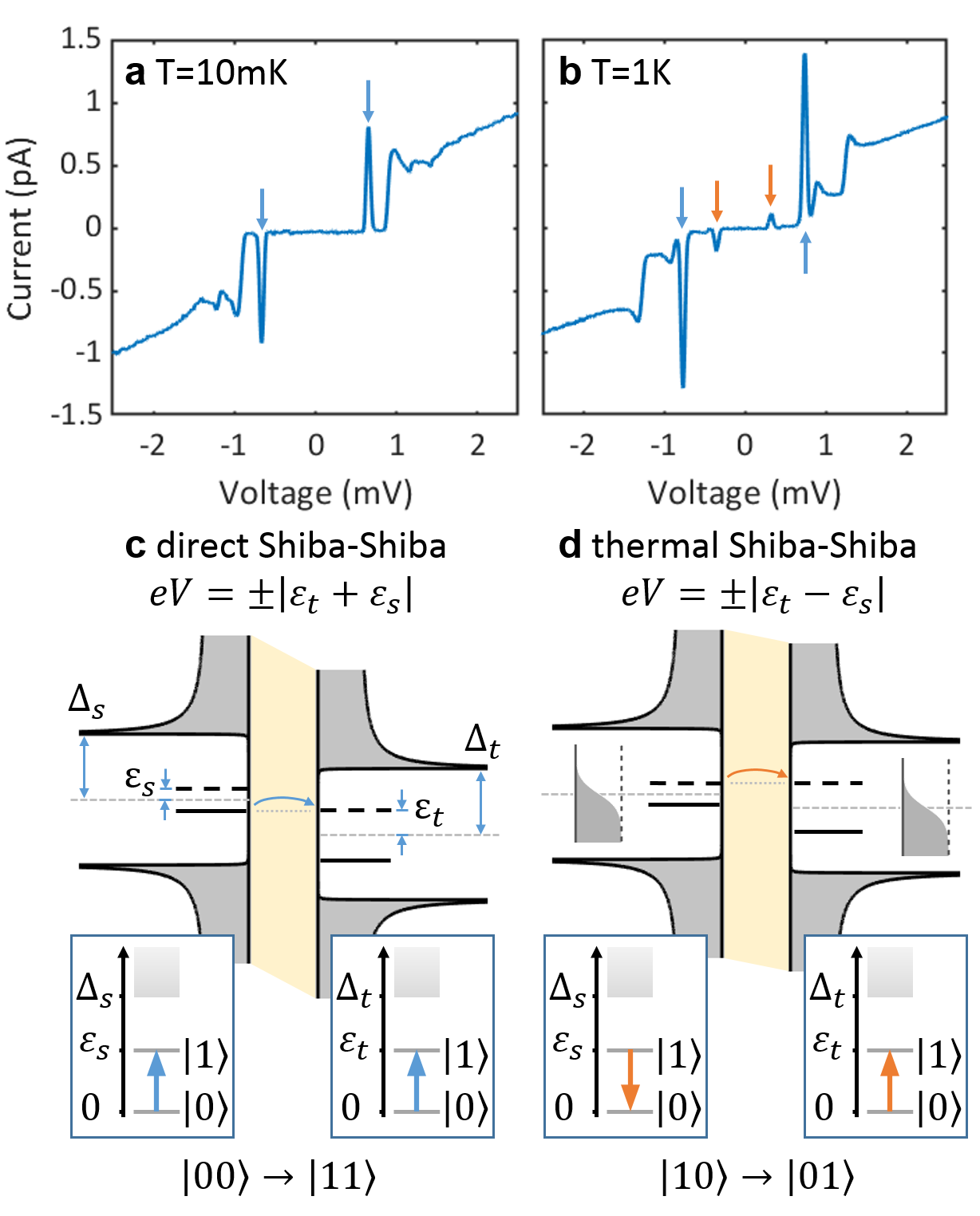}}
\caption{\textbf{Direct vs.\ thermal Shiba-Shiba tunneling.} \textbf{a,} $I(V)$ spectrum of Shiba-Shiba tunneling measured at 10\,mK. The blue arrows mark the peaks for direct Shiba-Shiba tunneling at $eV=\pm(\varepsilon_\textrm{t}+\varepsilon_\textrm{s})$. \textbf{b,} $I(V)$ spectrum of Shiba-Shiba tunneling measured at 1\,K. The red arrows mark the peaks for thermal Shiba-Shiba tunneling at $eV=\pm|\varepsilon_\textrm{t}-\varepsilon_\textrm{s}|$ (blue arrows: direct Shiba-Shiba peaks). \textbf{c,} Direct Shiba-Shiba tunneling process. The spectral functions in tip and sample are shifted by the bias voltage $eV=\varepsilon_\textrm{t}+\varepsilon_\textrm{s}$. The system starts in the ground state ($|00\rangle$). Tunneling of an electron leaves both tip and sample in an excited state ($|11\rangle$). This is illustrated in the energy diagrams (insets). \textbf{d,} Thermal Shiba-Shiba tunneling: The spectral functions in tip and sample are shifted by the bias voltage $eV=|\varepsilon_\textrm{t}-\varepsilon_\textrm{s}|$. The system starts in a thermally excited state ($|10\rangle$). Thermally activated tunneling effectively transfers a quasiparticle across the junction ($|01\rangle$). The grey dashed line denotes zero energy.} \label{fig:Fig2}
\end{figure}

If, however, we position a tip functionalized with a YSR state over a YSR state in the sample, the spectrum changes significantly (yellow line in Fig.\ \ref{fig:Fig1}d). In addition to the conventional YSR tunneling processes at the energy $\pm(\varepsilon_{\textrm{\small s,t}} + \Delta_{\textrm{\small t,s}})$, there is a feature at $\pm(\varepsilon_{\textrm{\small s}}+\varepsilon_{\textrm{\small t}})$ inside the shaded area with significant negative differential conductance (NDC). This additional feature can be directly associated with tunneling between the isolated YSR state in the tip and that in the sample without any contribution from the quasiparticle continuum \cite{Saldana2018}. The energy of this feature $|\varepsilon_\textrm{\small t}+\varepsilon_\textrm{\small s}|$ is always smaller than the energy needed to access the quasiparticle continuum ($\Delta_\textrm{\small t,s}+\varepsilon_\textrm{\small s,t}$). In the following, we will refer to this new tunneling process between two YSR states as \textit{Shiba-Shiba tunneling}. In the corresponding current spectrum, an isolated peak can be observed at this energy, which is shown in Fig.\ \ref{fig:Fig2}a (blue arrows). Having zero current on either side of this peak is a clear signature of tunneling between single levels protected by a gap \cite{VanderVaart1995,Fujisawa1998}. These observations are robust across many Shiba-Shiba systems and do not rely on a specific value of the parameters $\Delta_\textrm{\small t,s}$ or $\varepsilon_\textrm{\small t,s}$.

\section*{Direct and Thermal Shiba-Shiba Processes}

At low temperature (10\,mK), the $I(V)$ curve in Fig.\ \ref{fig:Fig2}a shows one pair of Shiba-Shiba peaks located at a bias voltage of $eV=\pm|\varepsilon_\textrm{\small t}+\varepsilon_\textrm{\small s}|$ (blue arrows), which we refer to as \textit{direct Shiba-Shiba tunneling}. The process is explained schematically in Fig.\ \ref{fig:Fig2}c, where the spectral functions of tip and sample are shifted by the bias voltage $V$, such that direct Shiba-Shiba tunneling occurs when the filled level (solid line) of the sample YSR state is aligned with the unfilled level (dashed line) of the tip YSR state \cite{Saldana2018}. At high temperature (1\,K), the $I(V)$ curve in Fig.\ \ref{fig:Fig2}b features another pair of peaks located at a bias voltage of $eV = \pm |\varepsilon_{\textrm{\small s}}-\varepsilon_{\textrm{\small t}}|$ (red arrows), clearly distinct from the direct Shiba-Shiba process (blue arrows). Looking at the aligned spectral functions in Fig.\ \ref{fig:Fig2}d, the YSR level above the Fermi level (grey dashed line) will be thermally occupied, such that tunneling to the empty YSR level (dashed line) on the other side is possible. Therefore, in the following we refer to this thermally activated process as \textit{thermal Shiba-Shiba tunneling}.

Looking beyond the tunneling process at what happens inside the superconductor, the simple picture of direct Shiba-Shiba tunneling outlined above has to be adapted to the fact that the elementary excitations in a superconductor are Bogoliubov quasiparticles (a superposition of electrons and holes), but only electrons or holes can tunnel. As a result, after the direct Shiba-Shiba tunneling process  Bogoliubov quasiparticles are excited in both tip and sample conserving even parity, which is schematically shown in Fig.\ \ref{fig:Fig2}c. In Dirac notation, the tunneling process is expressed by $|00\rangle\rightarrow|11\rangle$, where $|s,t\rangle$ describes the state of sample and tip with $s,t = 0,1$ denoting the ground and excited state, respectively (for details see the Supplementary Information \cite{supinf}). Both excited Bogoliubov quasiparticles need to subsequently relax to recover the original situation ready for the next tunnel event.

The thermal Shiba-Shiba process, on the other hand, resembles a conventional tunneling process (Fig.\ \ref{fig:Fig2}d), i.\ e.\ a particle is destroyed in one side and created in the other conserving odd parity. In Dirac notation, we write $|10\rangle\rightarrow|01\rangle$. The net current for the thermal Shiba-Shiba process is significantly reduced due to the small probabilities for the YSR states to be thermally occupied. At 1\,K, this reduction is on the order of a few percent for typical YSR energies in our system, which makes the thermal Shiba-Shiba peaks clearly visible in the spectrum. At a temperature of 10\,mK, we fully suppress the thermal Shiba-Shiba process because of the exponential dependence of thermal excitation on temperature. In the following, we will focus on direct Shiba-Shiba tunneling processes at 10\,mK, unless stated otherwise.

\section*{Minimizing Relaxation Channels}

\begin{figure}
\centerline{\includegraphics[width = \columnwidth]{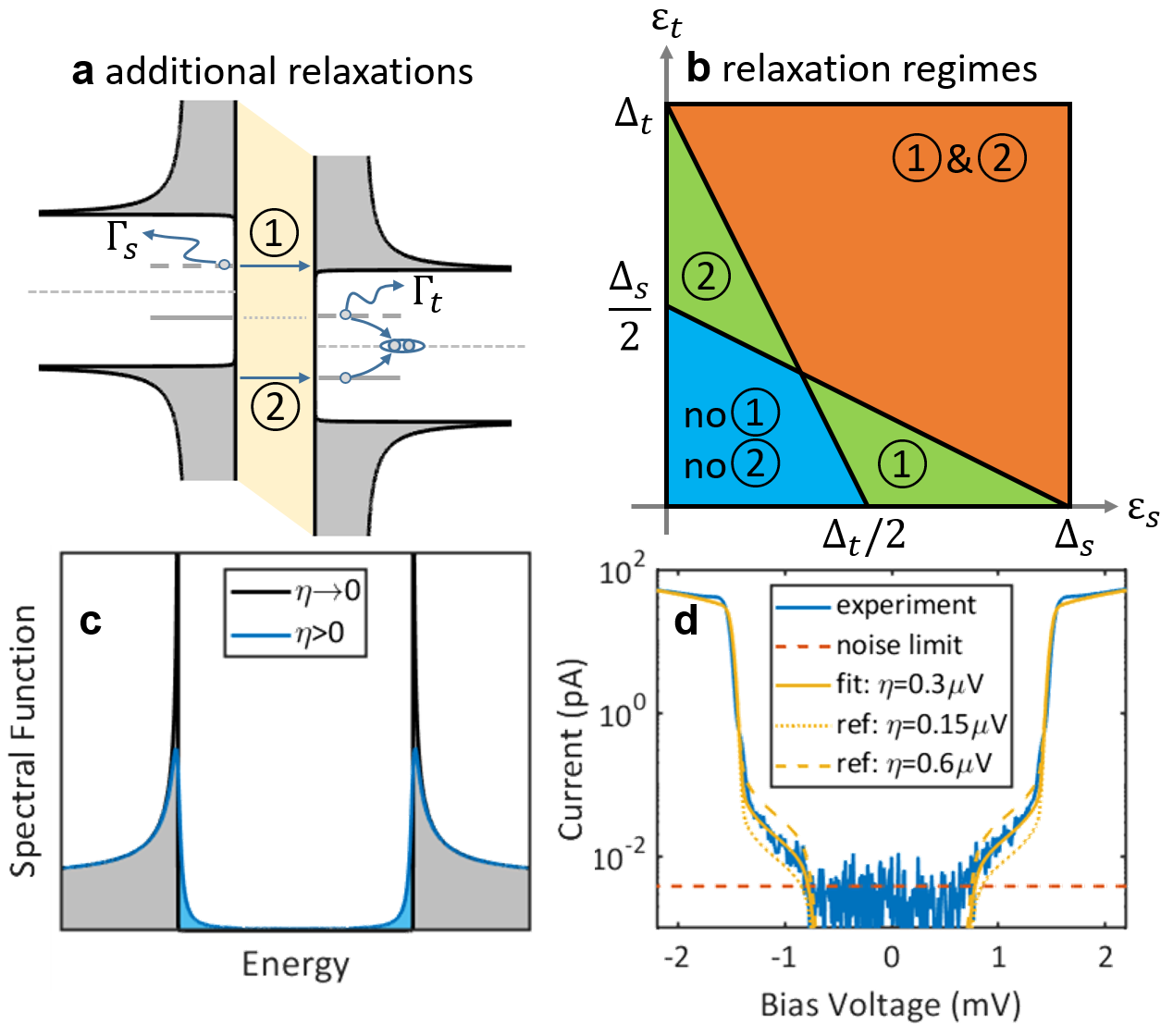}}
\caption{\textbf{Relaxation Channels and Regimes.} \textbf{a,} Relaxation processes for the excited YSR states. The wavy arrows ($\Gamma_\textrm{\small s,t}$) describe intrinsic, parity breaking relaxation processes. In process 1, the excited YSR state on the left side is relaxed by tunneling into the continuum of the right side. In process 2, the YSR state on the right side relaxes by forming a Cooper pair with a quasiparticle that has tunneled from the continuum in the left side. \textbf{b,} Relaxation phase diagram. In the blue regime, relaxation has to happen through the intrinsic processes $ \Gamma_\textrm{\small s,t}$. \textbf{c,} Illustration of Dynes' $\eta$ as a phenomenological broadening parameter. \textbf{d,} Direct measurement of remnant quasiparticle gap filling with a clean tip and clean sample (current is absolute value plotted on a logarithmic scale to reveal a small but finite signal inside the gap). The detection limit of a few fA is indicated by a dashed horizontal line.} \label{fig:relax}
\end{figure}

Our experimental data clearly confirms tunneling between two individual quasiparticle levels. We, therefore, demonstrate the absolute limit of transport both in space and in energy. The reduction of the constituents for tunneling to a bare minimum turns the tunneling dynamics to a purely sequential process governed by the lifetime of the excited YSR states. Depending on the energy of the YSR states in the tip and the sample, different channels may be available for relaxing into the ground state. If the energies of the YSR states are close to zero energy as in Fig.\ \ref{fig:Fig2}c, only intrinsic relaxation channels are available (excess quasiparticles). If the energies of the YSR states are close to the gap edge, additional relaxation channels open up at higher conductance through tunneling to and from the continuum of the other electrode, which are indicated as process 1 and 2 in Fig.\ \ref{fig:relax}a. The intrinsic relaxation channels are indicated by the lifetime broadening parameter $\Gamma_\textrm{\small s,t}$. A diagram indicating the occurrence of the different relaxation processes as function of the YSR state energies is shown in Fig.\ \ref{fig:relax}b. In conventional YSR tunneling, process 1 and 2 always occur, which has been discussed in the context of resonant Andreev tunneling before \cite{Ruby2015a}. The unique aspect of Shiba-Shiba tunneling is that these relaxation channels can be eliminated so that the possible relaxation channels are reduced to a bare minimum in the blue region in Fig.\ \ref{fig:relax}b. In the following, we will focus on Shiba-Shiba tunneling in this blue region.

\section*{Extracting the Intrinsic Lifetime}

\begin{figure}
\centerline{\includegraphics[width = \columnwidth]{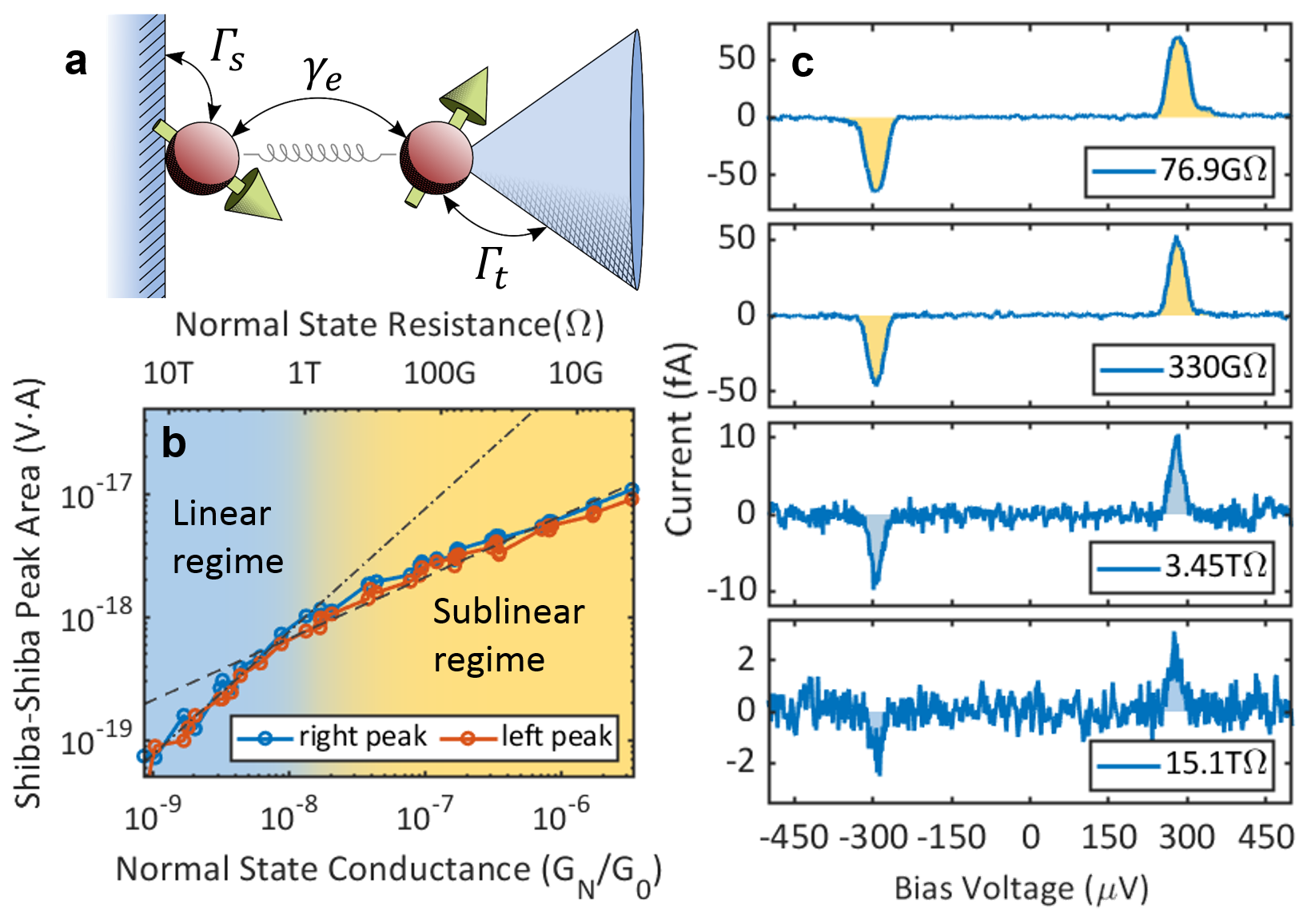}}
\caption{\textbf{Tunneling Dynamics.} \textbf{a,} Effective picture of the tunneling and relaxation processes, which shows the relation of the tunnel coupling $\gamma_e$ to the relaxation channels $\Gamma_\textrm{\small s,t}$. \textbf{b,} Conductance dependency of direct Shiba-Shiba current peak area. The linear to sublinear transition is clearly visible at around $10^{-8}\,G_0$. The lines indicate linear (dash-dot) and square root (dashed) behavior. \textbf{c,} The evolution of the direct Shiba-Shiba current peak with different normal state tunneling conductances. The upper two panels are in the sublinear regime, while the lower two panels are in the linear regime. The lowest panel is measured close to the current detection limit of $\approx 0.5\,$fA.} \label{fig:Fig3}
\end{figure}

The tunneling dynamics in the blue region is governed by the lifetime broadening in tip and sample $\Gamma_\textrm{\small s,t}$ as well as the tunnel coupling $\gamma_e$, which is schematically shown in Fig.\ \ref{fig:Fig3}a. Due to the sequential nature of the tunneling process, the system needs time to relax before the next tunneling event is possible. This condition holds as long as the tunnel coupling is small, i.\ e.\ $\gamma_e\ll \Gamma_\textrm{\small t,s}$. For strong tunnel coupling, i.\ e.\ $\gamma_e\gg \Gamma_\textrm{\small t,s}$, chances increase for higher order multiple tunneling (back and forth) between the YSR states without transferring additional charge during the process, which blocks the junction and reduces the current. The tunnel coupling is experimentally accessible through the setpoint conductance $G_\textrm{\small N}\propto\gamma_e^2$ \cite{supinf}. With the unique possibility to tune the tunnel coupling by varying the tip-sample distance, we can directly access the crossover between these two regimes.

The tunneling dynamics is reflected in the Shiba-Shiba peak area, which is shown in Fig.\ \ref{fig:Fig3}b as function of $G_\textrm{\small N}$. The peak area first scales linearly (enough time to relax) followed by a sublinear regime with a square root dependence (not enough time to relax) as function of setpoint conductance $G_\textrm{\small N}$. The transition into the sublinear regime describes an emergent coherent interaction between the excited YSR states within their lifetime. This could provide a path towards entanglement of the two Bogoliubov quasiparticles filling the excited YSR levels, which are separated in real space by the tip-sample distance. The extremely low lying transition point is indicative of a comparatively long lifetime.

This linear to sublinear transition can also be seen directly in the raw data of the Shiba-Shiba peak shown in Fig.\ \ref{fig:Fig3}c. In the lower two panels the peak current scales linearly with the tunneling resistance, but only at extremely high tunneling resistances in the Teraohm regime. In the upper two panels, the scaling is sublinear. Despite the small values of the current, tunneling is actually quite efficient. We would need a roughly 100 times higher bias voltage of about 30\,mV to get a comparable current in the normal conducting regime. This is due to the YSR spectral function being effectively a singularity in the limit of infinite lifetime. Considering the peak reduction due to finite lifetime as well as the interaction with the environment shows how efficient this sequential tunneling process actually is.

Going beyond the intuitive explanation above, we provide a full theory describing Shiba-Shiba tunneling, the details of which can be found in the Supplementary Information \cite{supinf}. The central result of this theory is the derivation of the Shiba-Shiba current
\begin{equation}
I_0(v)=\frac{e}{\hbar}\frac{\Gamma\gamma_e^2}{4\gamma_e^2+\Gamma^2+v^2} \label{eq:I0},
\end{equation}
which has Lorentzian shape. For simplicity, we assume equal lifetime broadening $\Gamma$ for the YSR states in tip and sample ($\Gamma=\Gamma_\textrm{\small s}=\Gamma_\textrm{\small t}$). The full expression for $\Gamma_\textrm{\small s}\neq\Gamma_\textrm{\small t}$ can be found in the Supplementary Information \cite{supinf}. The voltage $v$ is the detuning of the bias voltage $V$ from the level alignment $\varepsilon_\textrm{\small s}+\varepsilon_\textrm{\small t}$, i.\ e.\ $eV = \varepsilon_\textrm{\small s}+\varepsilon_\textrm{\small t} + ev$. The area under the Shiba-Shiba peak is
\begin{equation}
A=\frac{\pi\Gamma}{\hbar}\frac{\gamma_e^2}{\sqrt{4\gamma_e^2+\Gamma^2}},
\label{eq:area}
\end{equation}
which presents an important quantity that only becomes well-defined due to the discrete quasiparticle level tunneling. Reliably extracting the quasiparticle lifetime from any other quantity (e.\ g.\ height or width) is extremely difficult, because the peak shape changes due to external noise as well as the dissipative interaction with the electromagnetic environment as described by $P(E)$-theory \cite{Devoret1990,Averin1990,Ingold1992,Ingold1994,Fujisawa1998,Ast2016,supinf}. Since the $P(E)$ function is a probability function with normalized area, the area of the Shiba-Shiba peak does not change after convolution with the environmental interaction. The evolution of the Shiba-Shiba peak area in Eq.\ \ref{eq:area} agrees well with the experiment (see Fig.\ \ref{fig:Fig3}b); in the linear regime $A\propto \gamma_e^2\propto G_\textrm{\small N}$ (dash-dotted line) and in the sublinear regime $A\propto \gamma_e\propto \sqrt{G_\textrm{\small N}}$ (dashed line). At the linear to sublinear transition where $2\gamma_e = \Gamma$, the peak area is given by $A_\textrm{\small trans} = \pi\Gamma^2/(4\sqrt{2}\hbar)$, which gives direct experimental access to the lifetime $\Gamma$. For the data in Fig.\ \ref{fig:Fig3}b, the peak area of about $A_\textrm{\small trans} = 1\times 10^{-18}\,$VA translates to a combined lifetime broadening of $\Gamma=0.1\,\mu$eV, corresponding to a lifetime of $\tau = h/\Gamma = 41$\,ns. As we have minimized the relaxation channels in this system (cf.\ blue region in Fig.\ \ref{fig:relax}b), we conclude that the extracted lifetime is the intrinsic lifetime of the YSR states due to excess quasiparticles. If the superconducting substrate is further optimized for the reduction of residual quasiparticles, the lifetime would even be longer. In such a situation, lifetimes in excess of 100\,$\mu$s have been reported \cite{zgirski_evidence_2011,olivares_dynamics_2014}.

\section*{Residual Relaxation Mechanisms}

Looking at the intrinsic lifetime limiting mechanisms in more detail, phonons are excluded due to their exponential suppression at very low temperatures ($T\ll 1\,$K) \cite{Kozorezov2008,supinf}. The only feasible relaxation channel is through excess quasiparticles, which recombine with the excited Bogoliubov quasiparticles \cite{zgirski_evidence_2011,olivares_dynamics_2014,Martin2014}. This is schematically shown in Fig.\ \ref{fig:relax}c, where we model the effect of excess quasiparticles by adding a phenomenological broadening parameter $\eta$ \cite{Dynes1978}. Interestingly, this parameter introduces a finite parity lifetime translating directly into a finite YSR state broadening, even in the simplest YSR model \cite{Salkola1997,Flatte1997,Balatsky2006}. Excess quasiparticles can occur even at lowest temperatures due to, e.\ g., a paramagnetic background of defects or imperfect crystallinity \cite{Feldman2017}. We can directly observe this phenomenon in the current spectrum measured with a clean vanadium tip on a clean spot on the sample (Fig.\ \ref{fig:relax}d), and fit this spectrum with a broadening parameter $\eta = \eta_\textrm{\small t,s}=0.3\,\mu$eV (yellow line). To demonstrate the sensitivity of the model we have plotted the same spectra with half and double the broadening (yellow dotted and dashed lines), which show significant deviations. Based on this analysis, the lifetime broadening of the excited YSR state translates to $\Gamma=0.3\,\mu$V \cite{supinf}, which is the same order of magnitude of what we measured from Shiba-Shiba tunneling experimentally. Although this phenomenological approach cannot provide a detailed explanation of the relaxation mechanism, we conclude that excess quasiparticles could be a viable intrinsic relaxation channel for the YSR states at low temperatures (10\,mK).

\section*{Conclusion and Outlook}
Shiba-Shiba tunneling is a realization of tunneling between single quasiparticle levels at the atomic scale. It constitutes the ultimate limit in energy and space of a transport current through a tunnel junction. The tunneling dynamics of such a junction reveals the lifetime of individual levels protected by the superconducting gap at the atomic scale. The relaxation mechanism revealed in connection to in-gap quasiparticle background gives a prospect to further develop long lived single quasiparticle states.

Going beyond this proof-of-principle, this minimal configuration can be exploited for time dependent manipulation of a YSR state through parity conserving excitation during a tunneling event for both even ($|00\rangle\rightarrow|11\rangle$) and odd ($|10\rangle\rightarrow|01\rangle$) parity (at different bias voltages). Parity breaking mechanisms, such as excess quasiparticles can be reduced to increase lifetime or deliberately introduced to operate on the YSR states, e.\ g.\ through microwaves \cite{janvier2015}. As such, we have demonstrated a tunneling process at the atomic scale that provides great potential for a deeper understanding of engineering and manipulating superconducting bound states. And by extension, this provides a potential path towards detection and manipulation of Majorana bound states by tunneling from YSR states.

\section*{Methods}
The experiments were carried out in a scanning tunneling microscope (STM) operating at a base temperature of 10\,mK \cite{Assig2013}. The sample was an V(100) single crystal with $>99.99\%$ purity. To obtain a clean surface, the sample was prepared by multiple cycles of argon ion sputtering at $10^{-7}\,$mbar to $10^{-6}\,$mbar argon pressure with about 1\,keV acceleration energy and annealing at around $700^\circ$C. The sample is heated up and cooled down slowly to reduce strain in the crystal. The tip material was a polycrystalline vanadium wire of 99.8\% purity, which was cut in air and prepared in ultrahigh vacuum by sputtering and field emission.

The so obtained V(100) crystal exhibits a sparse concentration of intrinsic impurities at the surface, giving rise to YSR states. The origin of these magnetic impurities is probably a complex involving an oxygen vacancy, which is discussed in detail in the Supplementary Information \cite{supinf}. YSR states on the tip were created by controlled tip indentation implemented in a LabVIEW program, enabling automated and reliable fabrication of desired YSR state properties. See Supplementary Information \cite{supinf} for more details.

For measurements detecting currents above around 20\,fA, a Femto DLPCA-200 was used. For smaller currents down to 1\,fA, we use a Keysight B2983A instead. Within the overlap range between the two devices, the measurements on the same Shiba-Shiba area show consistency. The bias voltage is supplied by one output channel of a Nanonis SPM controller, attenuated by a 1:100 passive voltage divider before applied on the sample. The differential conductance (\textit{dI/dV}) curves shown in Fig.\ \ref{fig:Fig1}d were measured using standard lock-in techniques with a modulation of 10$\,\mu$V at 392.348\,Hz. STM/STS data were analysed and plotted using self-written code in MATLAB.

The dynamics between YSR levels is described by a coupled two level system. The resulting Shiba-Shiba current is calculated with a quantum master equation of the density matrix (Eq. \ref{eq:area}). For more details see the Supplementary Information \cite{supinf}.

\section*{Acknowledgments}
This work was funded in part by the ERC Consolidator Grant AbsoluteSpin (Grant No.\ 681164) and by the Center for Integrated Quantum Science and Technology (IQ$^\textrm{\small ST}$). J.A. acknowledges funding from the DFG under grant number AN336/11-1. A.L.Y. and J.C.C. acknowledge funding from the Spanish MINECO (Grant No. FIS2017-84057-P and FIS2017-84860-R), from the “Mar\'{\i}a de Maeztu” Programme for Units of Excellence in R\&D (MDM-2014-0377).

\newpage
\onecolumngrid
\begin{center}
\textbf{\large Supplementary Information}
\strut
\vspace{1em}
\end{center}
\setcounter{figure}{0}
\setcounter{table}{0}
\setcounter{equation}{0}
\renewcommand{\thefigure}{S\arabic{figure}}
\renewcommand{\thetable}{S\Roman{table}}
\renewcommand{\theequation}{S\arabic{equation}}
\twocolumngrid

\subsection*{Tip and sample preparation}

The sample was an V(100) single crystal with $>99.99\%$ purity, and was prepared by standard UHV metal preparation procedure of multiple cycles of argon ion sputtering at $10^{-7}\,$mbar to $10^{-6}\,$mbar argon pressure with about 1\,keV acceleration energy and annealing at around $700^\circ$C. The sample is heated up and cooled down slowly, around $1^\circ$C/s, to reduce strain in the crystal. The tip material was a polycrystalline vanadium wire of 99.8\% purity, which was cut in air and prepared in ultrahigh vacuum by Argon sputtering. Subsequent field emission on V(100) surface as well as standard tip shaping techniques were used to obtain a tip exhibiting clean bulk gap as well as good imaging capabilities.

\subsection*{V(100) surface and defects}

\begin{figure}
\centerline{\includegraphics[width = \columnwidth]{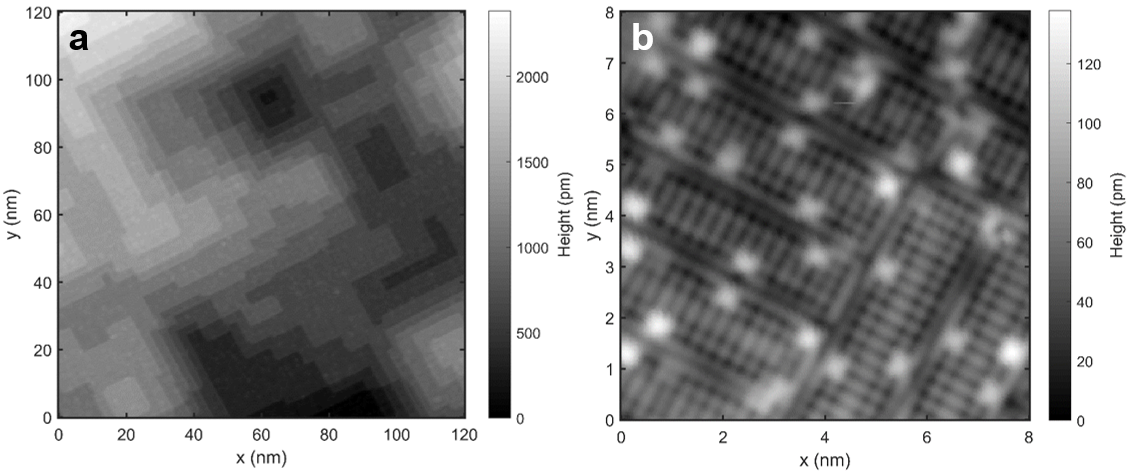}}
\caption{\textbf{(a)} Atomic terraces on the V(100) surface imaged with 50\,mV and 50\,pA. \textbf{(b)} Topographic image measured at 4\,mV and 1\,nA showing the $(5\times 1)$ oxygen reconstruction of V(100) surface with oxygen vacancies as bright protrusions.
} \label{fig:Fig_si_1}
\end{figure}

Vanadium is a conventional Bardeen-Cooper-Schrieffer (BCS) superconductor, with a transition temperature $T_\text{C}=5.4\,$K and an order parameter $\Delta_\text{t,s}=750\pm 10\,\upmu$eV for both tip and sample. At a base temperature of 10\,mK both tip and sample are well superconducting. The V(100) surface typically shows a $(5\times 1)$ reconstruction due to the diffusion of bulk oxygen impurities to the surface through sputtering-annealing cycles during sample preparation \cite{si_Jensen1982,si_Koller2001,si_Dulot2001}. In an STM topography, the V(100) surface shows typical atomic terraces (Fig.\ \ref{fig:Fig_si_1}(a) indicating crystalline nature of the substrate. A zoom-in image shows atomic resolution (Fig. \ref{fig:Fig_si_1}(b)), where the ($5\times 1$) reconstruction induced by diffused oxygen from the bulk can be clearly seen \cite{si_Jensen1982,si_Koller2001,si_Kralj2003}. We not only see ($5\times 1$), but also ($6\times 1$) or ($4\times 1$) reconstruction with boundaries, which was attributed to a local change in chemical environment \cite{si_Kralj2003}.

\begin{figure}
\centerline{\includegraphics[width = \columnwidth]{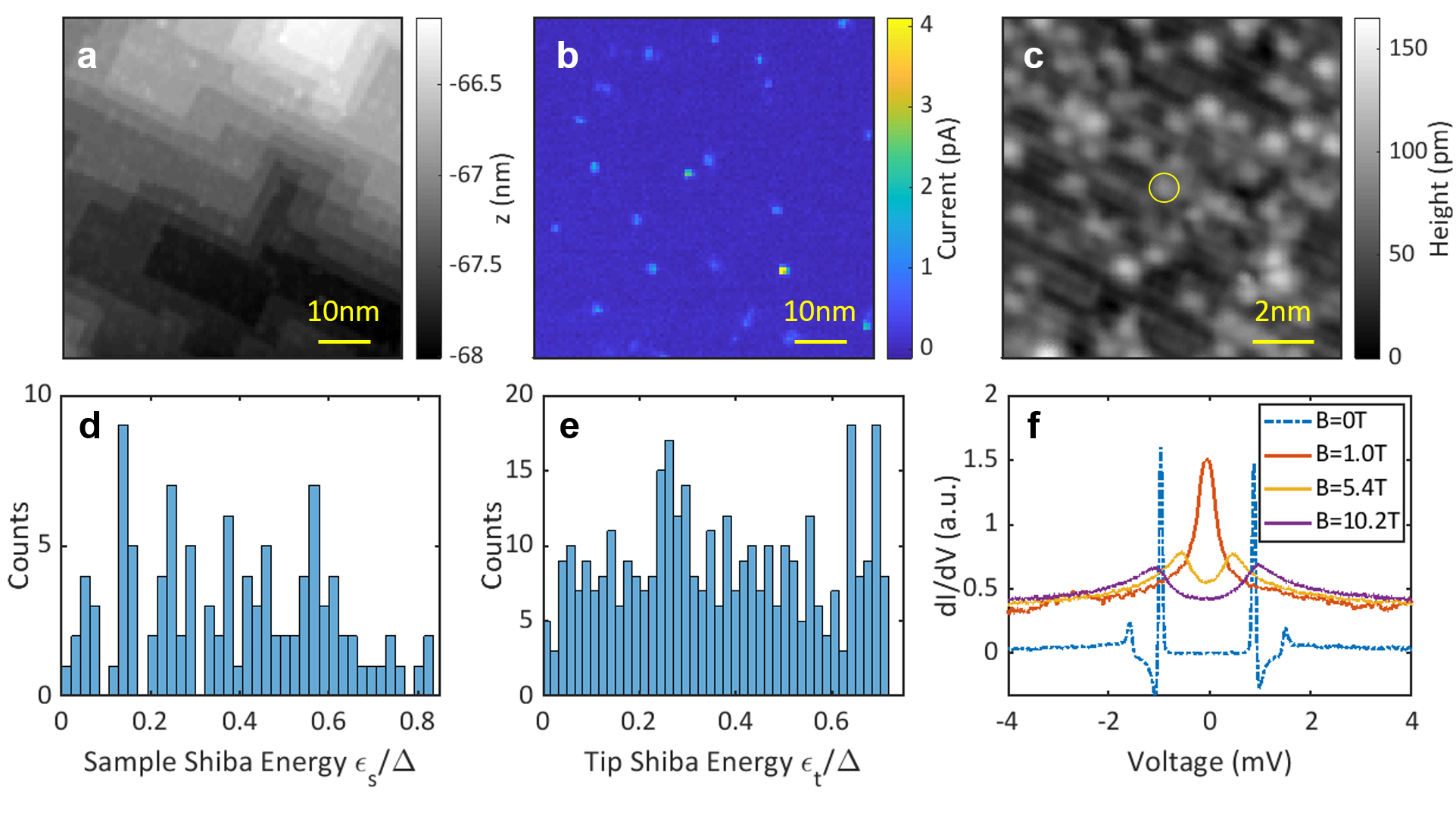}}
\caption{\textbf{(a)} Topographic image of the V(100) surface at 4\,mV and 10\,pA. \textbf{(b)} Local current measurement with spatial resolution at a bias voltage of 1.42\,mV, which is just inside the superconducting gap. The bright spots indicate defect positions with YSR states on the surface. \textbf{(c)} Atomically resolved image near a YSR defect (labelled with a circle) at 4\,mV 100\,pA. Other bright protrusions are possibly oxygen vacancies. \textbf{(d)} Histogram of the energy distribution of intrinsic YSR states on V(100) surface. \textbf{(e)} Histogram of the energy distribution of YSR states introduced on the V tip. \textbf{(f)} Bias spectra of a YSR tip in a magnetic field showing the evolution from YSR state at zero field to Kondo splitting at high field.} \label{fig:Fig_si_2}
\end{figure}

Besides the periodic reconstruction structure, there is a finite concentration of defects on the surface. The most abundant ones that are visible under our imaging conditions (at various set points) appears to be small bright protrusions of about 50\,pm in height as can be seen in Fig. \ref{fig:Fig_si_2}. The concentration is estimated to be around 4\% of a monolayer (ML) varying somewhat depending on the sample preparation procedure. These defects are likely to be oxygen vacancies \cite{si_Bergermayer2002,si_Bischoff2002}, since electronegative atoms such as oxygen or carbon atoms typically deplete the density of states around and appear dark in the topography \cite{si_Bischoff2001}. Most of the oxygen vacancies do not show any magnetic signature like a Yu-Shiba-Rusinov (YSR) state or a Kondo peak, so we conclude that an oxygen vacancy alone is likely not the origin of the intrinsic YSR defects discussed in the main text.

In addition to oxygen vacancies, we may have a very low carbon concentration on the surface, because of the slow cooling process after annealing, which results in carbon atom segregation to the surface \cite{si_Bischoff2002}. Unfortunately, in the STM topography, it is not possible to clearly distinguish carbon from oxygen, since carbon can freely substitute oxygen atoms in the fourfold hollow sites \cite{si_Bischoff2002} and the apparent height difference is small\cite{si_Bergermayer2002}. However, the fact that we observe different superstructures besides ($5\times 1$) and that random but small height variation within the $(5\times 1)$ reconstruction, suggests a locally varying carbon concentration and widespread existence of carbon \cite{si_Kralj2003}. These various structures typically do not show any YSR state, indicating that carbon atoms alone do not have spin signature.

Apart from these two major categories of non-magnetic defects on the surface (oxygen vacancies and carbon substitutions), there is a small amount of magnetic defects that exhibit YSR states. To find those YSR defects, we measure various $I(V)$ spectra on a spatial grid and search for states inside the superconducting gap. Fig.\ \ref{fig:Fig_si_2}(a) shows the surface topography taken in such a typical measurement with a current map at a voltage just below the gap edge at 1.42\,mV shown in Fig.\ \ref{fig:Fig_si_2}(b). The coherence peak at the gap edge is at 1.52\,mV. A YSR state will appear as a step in the current inside the gap, which will leave a non-zero current at 1.42\,mV. The tip is clean and superconducting, such that a non-zero current denotes an in-gap state indicating a YSR state. A zoom-in of the topography in Fig.\ \ref{fig:Fig_si_2}(c), where the impurity exhibiting a YSR state is labeled with a yellow circle. Note that there is only one defect with a YSR state in this topography image.

By analyzing the occurrence of YSR states in the grid spectra, we find the concentration of impurities with YSR states to be around $0.02\%-0.05\%$ of a ML. However, due to the finite spatial extension of YSR states ($1-2$\,nm), sub-surface defects may also be detected and counted, and thus the above estimation overestimates the real concentration. Since foreign atoms with magnetic properties like other transition metals are not reported to segregate to the V(100) surface, we argue that those defects showing YSR states originate from a complex combination of simple elements with unpaired spin, for example a carbon-oxygen vacancy complex. The fact that those intrinsic defects with YSR states usually show only one pair of YSR peaks (usually there will be multiple YSR states for transition metal atoms) and that we can introduce YSR states to the tip reliably and reproducibly by random dipping of the tip (it is statistically difficult to coincidentally pick up a magnetic impurity whose concentration is already very rare) supports this claim.

\subsection*{Intrinsic YSR defects}
Despite their origin and dilute concentration, the intrinsic YSR states on the V(100) surface have many desired features that surpass those of deposited atoms or molecules. They usually feature only one pair of YSR states inside the gap. YSR states generated from deposited atoms typically feature multiple pairs when the resolution of the setup is high enough to resolve them, such that the cross-relaxation between them requires advanced modelling \cite{si_Ruby2015a}. Here, the intrinsic defects featuring only one pair of YSR states are necessary for an unambiguous separation of Shiba-Shiba peaks and a lifetime estimation without any cross talk between multiple YSR states.

In addition, the intrinsic YSR defects have a wide energy distribution (Fig.\ \ref{fig:Fig_si_2}(d)), which means that a YSR state with almost any energy within gap can be found. This offers an extremely high design flexibility, in contrast to conventional deposited single magnetic atoms or molecules, where the small number of adsorption sites limits the range of YSR state energies. This is crucial, since as shown in the main text, there are different relaxation regimes depending on the YSR state energy.  Moreover, Shiba-Shiba tunneling is a general process regardless of the origin of the YSR state. Given the above advantages, we use the intrinsic YSR states in vanadium.

\subsection*{YSR tip}
For the complete Shiba-Shiba tunneling configuration, we need to introduce a YSR state to the apex of the tip. This can be done reliably by randomly dipping the tip into the vanadium surface. This process can be automated: the tip is first dipped with the voltage and depth at the position specified by the user, and then it is moved to a clean area where a bias spectroscopy spectrum is taken, after which the program checks whether there is a YSR state in the spectrum which satisfies certain criteria specified by the user. In this way, a tip with the desired YSR state at an arbitrary energy inside the gap, intensity and asymmetry can be controllably designed. A histogram of YSR state energies shown in Fig.\ \ref{fig:Fig_si_2}(e) demonstrates that the YSR state on the tip has total flexibility concerning its energy.

We can further prove the existence of a magnetic signature on the tip by applying a magnetic field to quench superconductivity and showing the replacement of the YSR state by a Kondo peak (Fig.\ \ref{fig:Fig_si_2}(f)). If the magnetic field is increased further, the Kondo peak splits with the splitting depending linearly on the applied field. This effect has been reported before \cite{si_Franke2011,si_Hatter2017} for magnetic defect on superconducting substrate, and here it proves that we do have a magnetic impurity on the apex of the tip.

It is worth mentioning a few concepts regarding the origin of magnetic defect on the tip. Statistically speaking, it is nearly impossible to pick up an intrinsic YSR defect randomly (or any specific element other than the major composition of the surface), because they are rare and the tip shaping condition is usually quite moderate that the crash site is merely about 5\,nm in diameter. In addition, the YSR state on the tip does not originate from the tip itself, e.\ g.\ some reconstruction or a simple rearrangement of certain defects at the apex. Rather, we find that tip indentation on the clean surface is much more efficient in making a tip with a YSR state than dipping the tip continuously on the same crash site, indicating a crucial role of migration of surface material to the tip. Consequently, the most probable candidates are oxygen and carbon in some special arraangement, which points to a similar origin as the intrinsic YSR states on the sample. This is supported by the lack of statistical difference of tip and sample YSR states (see Fig.\ \ref{fig:Fig_si_2}(d) and (e)).

\subsection*{Low conductance measurement}
The linear to sublinear transition happens at very low conductances $G_\text{N}$ for Shiba-Shiba tunneling in the blue regime (see Fig.\ 3(b) in the main text), typically with a setpoint current well below 100\,fA (at 4\,mV). We need a very low noise measurement of the current below 10\,fA to resolve the linear regime. The Femto DLPCA-200, which is the typical current-voltage amplifier we use, is best suited for currents above 20\,fA. For smaller currents, we use Keysight B2983A (picoammeter). Within the measurement range of 20\,pA, the picoammeter can detect currents below 1\,fA and is still fast enough to respond to the change of the bias voltage during a spectroscopy measurement. Consequently, we measure high conductance spectra with the Femto, which is faster in this range, and low conductance spectra with the picoammeter set to the 20\,pA measurement range. Within the overlap between the picoammeter and the Femto, the measurements on the same Shiba-Shiba area show consistency between the two devices.

Usually, the maximum Shiba-Shiba current is much higher than the setpoint current at low conductance $G_\text{N}$. Therefore, it is more difficult to measure the setpoint current than to measure Shiba-Shiba peak in spectrum. To circumvent this problem, we measure the setpoint current $I(z)$ at 4\,mV as a function of tip-sample distance $z$ with the Femto at high conductance $G_\text{N}$ and with the picoammeter at low conductance range. For the lowest conductance setpoints $G_\text{N}$, we extrapolate the exponential $I(z)$ dependency to extract the value for the tip-sample distance. The different methods have large overlapping ranges of applicability and they give consistent results when compared.

\subsection*{Transport theory on Shiba-Shiba tunneling}

\subsubsection*{Phenomenology}

In order to develop a theoretical model of tunneling between YSR states, we have to identify the dominant transport mechanism responsible for the observed narrow peak in the tunnel current. We argue that the dominant transport process must be composed of elementary transport events that carry a single charge $e$, rather than multiple charges. This will explain the large current peak observed at tunnel coupling values much smaller than those associated with the onset of Cooper pair tunneling (charge $2e$), or Andreev reflections (multiples of $e$).

When modeling charge transport in superconducting junctions, we have to account for the fact that the elementary excitations, Bogoliubov quasiparticles, are coherent superpositions of electrons and holes. In second quantization notation, the Bogoliubov quasiparticles are,
\begin{align}
    \gamma = & uc + v c^\dagger \\
    \gamma^\dagger = & vc + uc^\dagger
\end{align}
where $c^\dagger$, $c$ and $\gamma^\dagger$, $\gamma$ are the electron and Bogoliubov quasiparticle creation and annihilation operators, respectively. The coefficients $u$ and $v$ are the coherence factors for the electrons and holes. Here, we have suppressed the spin and momentum quantum numbers of the particle operators, with the implied convention that electronic operators $c^\dagger$ and $c$ act on time-reversed pairs of states (such as $\textbf{k}\uparrow$ and $-\textbf{k}\downarrow$).

Charge transport is described by the usual tunneling Hamiltonian that annihilates an electron in the sample and creates an electron in the tip $c_\text{s}c_\text{t}^\dagger$ (together with its reversed process $c_\text{t} c_\text{s}^\dagger$), which can be rewritten in terms of Bogoliubov quasiparticles,
\begin{equation}
    c_\text{s}c_\text{t}^\dagger = v_\text{s} u_\text{t}\, \gamma^\dagger_\text{s} \gamma^\dagger_\text{t}  - u_\text{s} v_\text{t}\, \gamma_\text{s} \gamma_\text{t} + u_\text{s} u_\text{t}\, \gamma_\text{s} \gamma^\dagger_\text{t} - v_\text{s} v_\text{t}\, \gamma^\dagger_\text{s} \gamma_\text{t}.
    \label{eq:tunnel}
\end{equation}
A similar equation can be found for the reversed process $c_\text{t}c_\text{s}^\dagger$. Because Bogoliubov quasiparticles (but not electrons and holes) populate well defined energy states, we can associate a change of energy $\delta E$ to each of the four quasiparticle operators,
\begin{align}
		\gamma^\dagger_\text{s} \gamma^\dagger_\text{t}:\ & \delta E = \varepsilon_\text{s} + \varepsilon_\text{t}\\
		\gamma_\text{s} \gamma_\text{t}:\ & \delta E = -(\varepsilon_\text{s} + \varepsilon_\text{t})\\
		\gamma_\text{s} \gamma^\dagger_\text{t}:\ & \delta E = \varepsilon_\text{t} - \varepsilon_\text{s}\\
		\gamma^\dagger_\text{s} \gamma_\text{t}:\ & \delta E = \varepsilon_\text{s} - \varepsilon_\text{t}
\end{align}
Each process described above can become resonant when the voltage bias provides the tunneling electron
with an energy matching $eV=\delta E$ for tunneling along the bias and $-eV=\delta E$ for tunneling against it.
Therefore, the processes $\gamma^\dagger_\text{s} \gamma^\dagger_\text{t}$ and $\gamma_\text{s} \gamma_\text{t}$
occuring at $eV=\pm(\varepsilon_\text{s} + \varepsilon_\text{t})$
are nicely separated in energy from processes $\gamma_\text{s} \gamma^\dagger_\text{t}$ and $\gamma^\dagger_\text{s} \gamma_\text{t}$ that
occur at $eV=\pm(\varepsilon_\text{s} - \varepsilon_\text{t})$.

At low temperatures (10\,mK), when $k_\text{B}T\ll \Delta$, both the bulk of the two superconductors (tip and sample), as well as the YSR states, are well described by their respective ground state. Therefore, no unpaired Bogoliubov quasiparticles are available for transport and processes $\gamma_\text{s} \gamma^\dagger_\text{t}$ and $\gamma^\dagger_\text{s} \gamma_\text{t}$ are suppressed. In this case, we expect resonances only at the voltages $eV=\pm(\varepsilon_\text{s} + \varepsilon_\text{t})$.

\begin{figure}
\begin{center}
\includegraphics[width=\columnwidth]{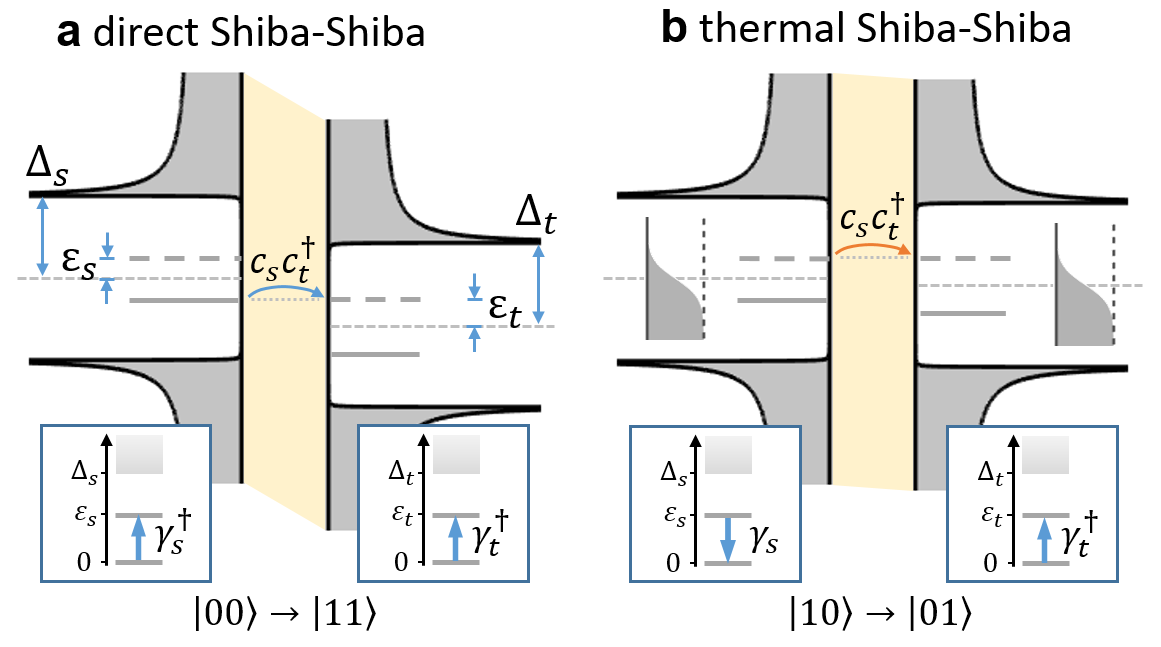}
\caption{\textbf{(a)} Direct Shiba-Shiba tunneling process. The operators indicate the corresponding process. After an electron is transferred in the tunneling event, the two superconductors are both left in an excited state. \textbf{(b)} Thermal Shiba-Shiba tunneling. This process is more like a conventional tunneling process, where a quasiparticle is destroyed on one side and created in the other.}
\label{fig:sketch}
\end{center}
\end{figure}

Fixing the voltage to $eV=\varepsilon_\text{s} + \varepsilon_\text{t}$, the term $v_\text{s} u_\text{t}\, \gamma^\dagger_\text{s} \gamma^\dagger_\text{t}$ in $c_\text{s}c_\text{t}^\dagger$ describes the process carrying a charge $e$ along the bias that creates two excited Bogoliubov quasiparticles. In Dirac notation, we write $|00\rangle\rightarrow|11\rangle$, as outlined in the main text.  This is illustrated in Fig.\ \ref{fig:sketch}. The other transport process relevant at the same voltage is the reverse process $|11\rangle\rightarrow|00\rangle$, accompanied by a charge $e$ tunneling backwards, against the bias.

The two processes describing tunneling of an electron back and forth across the barrier give rise to coherent Rabi oscillations between the ground state $|00\rangle$ and the excited YSR states $|11\rangle$.
Coherent oscillations on their own do not carry a dc current. However, energy relaxation processes will give rise to a resonance in the dc transport of Breit-Wigner type. This leads to the peak in the dc tunnel current. The peak width is given by the lifetime of the excited YSR states. The relaxation of a YSR state corresponds to the release of an excited quasiparticle into the continuum of states of the superconducting bulk. This process has a very long lifetime at low temperatures, as it must change the parity and, therefore, lift the protected symmetry of BCS superconductors. Our experiment, therefore, provides a direct tool to investigate the parity protection and its breakdown due to quasiparticle poisoning in a superconducting STM setup.

The Rabi frequency describing coherent oscillations between states $|00\rangle$ and $|11\rangle$ grows with the amplitude of the transport process. Therefore, as we increase the conductance by approaching the tip to the sample, the Rabi frequency will grow as well. Eventually, a threshold is reached when the Rabi frequency exceeds the rate of energy relaxation, giving rise to a coherent transport regime. The signature of emerging coherent transport is observed as a transition of the Shiba-Shiba peak area from a linear conductance dependence to a sublinear conductance dependence, as described in the main text and detailed below.

In the following, we describe a simple model that captures the described physical situation and allows us to calculate the amplitude of the leading order tunneling process, as well as the dc current peak as a function of the YSR states lifetime.

\subsubsection*{General Hamiltonian}

We split the Hamiltonian of the system into that describing the isolated tip and substrate $H_0$ and the tunnel Hamiltonian $H_\text{T}$ describing electrons hopping between the tip and substrate $H = H_0 + H_\text{T}$. In the following, we develop a perturbation theory in $H_\text{T}$. We denote by $H_\text{s}$ the Hamiltonian of the substrate and by $H_\text{t}$ the Hamiltonian of the tip, such that $H_0 = H_\text{s}+H_\text{t}$.  Hamiltonians $H_i$, with $i=\{\text{s},\text{t}\}$, are diagonal in the basis of quasiparticle states described by operators $\gamma_{ik\sigma}$.
The corresponding excitation energies, $\varepsilon_{ik\sigma}$, are described by the BCS continuum above the gap,
together with the in-gap YSR state $\varepsilon_i$.
\begin{align}
\displaystyle H_{i=s,t} = & \varepsilon_i \gamma^\dagger_{i,\uparrow}\gamma_{i,\uparrow} + \displaystyle\sum_{k\sigma} \varepsilon_{ik\sigma} \gamma^\dagger_{ik\sigma}\gamma_{ik\sigma}.
\end{align}
The tunnel Hamiltonian is given by,
\begin{align}
H_\text{T} = \sum_{k,k',\sigma} t\ c^\dagger_{sk\sigma} c_{tk'\sigma} + {\rm h.c.}
\end{align}
The approach is similar to Ref.\ \cite{si_Salkola1997,si_Ruby2015a}.

We are interested in the blue regime as presented in Fig.~2 of the main text. Therefore, we restrict the theoretical description to the situation when both YSR energies are sufficiently far from the edges of both gaps. In this case, the transport at voltages well below the gap does not involve the states in the continuum. We therefore focus exclusively on the subgap states of $H_0$. These are denoted $\ket{00}$, $\ket{01}$, $\ket{10}$, and $\ket{11}$, with $\ket{n_\text{s},n_\text{t}}$ being the state with $n_\text{s}$ quasiparticles occupying the excited YSR state in the substrate, and similarly $n_\text{t}$ for the excitations of the YSR state in the tip.

The YSR states can be described using the well known classical impurity spin model outlined in Ref. \cite{si_Balatsky2006}. In this case, the YSR energies are parametrized by two dimensionless quantities $\alpha_i=\pi \nu_i J_i S_i$ and $\beta_i = \pi \nu_i V_i$, for $i=\{\text{s},\text{t}\}$,
\begin{align}
    \varepsilon_i = \left| \Delta_i \frac{1-\alpha_i^2+\beta_i^2}{\sqrt{(1-\alpha_i^2+\beta_i^2)^2+4\alpha_i^2}}\right|,
\end{align}
where $\nu_i$ is the density of states, $J_i$ is the exchange coupling, $S_i$ is the impurity spin, $V_i$ is the Coulomb scattering potential, and $\Delta_i$ is the superconducting order parameter. However, the theory presented here is not restricted to the classical impurity model, but relies solely on the presence of a single non-degenerate in-gap YSR state. A YSR state can accommodate a single quasiparticle with spin oriented parallel to the corresponding impurity spin $\vec{S}_i$ \cite{si_Salkola1997}.

At low temperature, $k_\text{B}T \ll \varepsilon_i,\Delta$, the tip and substrate lie in their ground state, i.\ e.\ in state $\ket{00}$.
The excited state $\ket{11}$ has a finite lifetime determined by the relaxation rate $\Gamma_\text{s,t}$ of each
YSR excitation in superconductor $i$. The process of relaxation does not conserve the number of particles in the sub-gap states.
It can be pictured as the annihilation of the excited YSR quasiparticle and creation of a quasiparticle in the continuum of states
of the same superconductor. In contrast to the localized YSR state quasiparticle, a quasiparticle in the continuum is delocalized
and will quickly leave the junction. Because the relaxation process requires absorption of energy of the order $\Delta_i$,
it is very slow at low temperatures, when it requires the pre-existence of a bulk (excess) quasiparticle in the vicinity of the junction. This process typically sets the slowest timescale in superconducting devices and explains the long lifetime of excited YSR states.

The excited state $\ket{11}$ relaxes into the singly-excited states, $\ket{01}$ by rate $\Gamma_\text{s}$, and $\ket{10}$ by $\Gamma_\text{t}$, respectively. The singly-excited states $\ket{10}$ and $\ket{01}$ decay into the ground state with the rates $\Gamma_\text{s,t}$, respectively.

\subsubsection*{Tunneling dynamics}

At voltages near resonance, $eV=\varepsilon_\text{s}+\varepsilon_\text{t}+ ev$ with a detuning of $ev\ll \varepsilon_\text{s,t}$, we combine the coherent dynamics between even states $\ket{00}$ and $\ket{11}$ induced by tunneling, with the incoherent relaxation processes in the individual YSR states. The coherent dynamics is described by a simple $2\times 2$-Hamiltonian in the space of even states $\ket{00}$ and $\ket{11}$,

\begin{align}
H_e = \begin{bmatrix}
-v/2 & \gamma_e \\
\gamma_e^* & v/2
\end{bmatrix},\quad \gamma_e=\sandwich{{00}}{H_\text{T}}{{11}},
\label{He}
\end{align}
where $\gamma_e$ is the corresponding tunnel coupling amplitude, linear in $H_\text{T}$.

We note that the dynamics between odd states $\ket{10}$ and $\ket{01}$ induced by tunneling becomes resonant at a different value of voltage, $eV_\text{odd}=\varepsilon_\text{t}-\varepsilon_\text{s}$, and can be safely neglected in the discussion of the peak at $eV_\text{even}=\varepsilon_\text{s}+\varepsilon_\text{t}$, unless either of the YSR states is close to the quantum phase transition point when the YSR energy vanishes.

The coherent dynamics governed by $H_e$ must be supplemented with incoherent relaxation processes described by the corresponding relaxation rates. Assuming that the system is in the blue regime (cf.\ Fig.\ 3(b) of the main text), the following quantum master equation governs the dynamics of the density matrix $\rho$ with entries $\rho_{xy}\equiv\sandwich{x}{\rho}{y}$ in the space of the four states $\{x,y\}\in\{\ket{00},\ket{01},\ket{10},\ket{11}\}$,
\begin{align}
\hbar\frac{d}{dt}\rho =  i [H_e,\rho] - & (\Gamma_\text{s}+\Gamma_\text{t}) P_{11} \ket{11}\bra{11} \\
                                 + & (\Gamma_\text{s} P_{11}-\Gamma_\text{t} P_{01}) \ket{01}\bra{01}\notag\\
                                 + & (\Gamma_\text{s} P_{11}-\Gamma_\text{s} P_{10}) \ket{10}\bra{10}\notag\\
								 + & (\Gamma_\text{s}P_{10}+\Gamma_\text{t}P_{01})\ket{00}\bra{00} \notag\\
								 - & \frac{(\Gamma_\text{s}+\Gamma_\text{t})}{2} \sum_{x \neq y}
                                     \left(\rho_{xy}\ket{x}\bra{y}+\text{h.c.}\right),\notag
\end{align}
where we introduced the notation $P_x=\rho_{xx}$ for the state populations. The terms in the last line represent the loss of coherence through the decay of off-diagonal terms $\rho_{xy}$ with $x\neq y$.

\begin{figure}[ht]
\begin{center}
\includegraphics[width=\columnwidth]{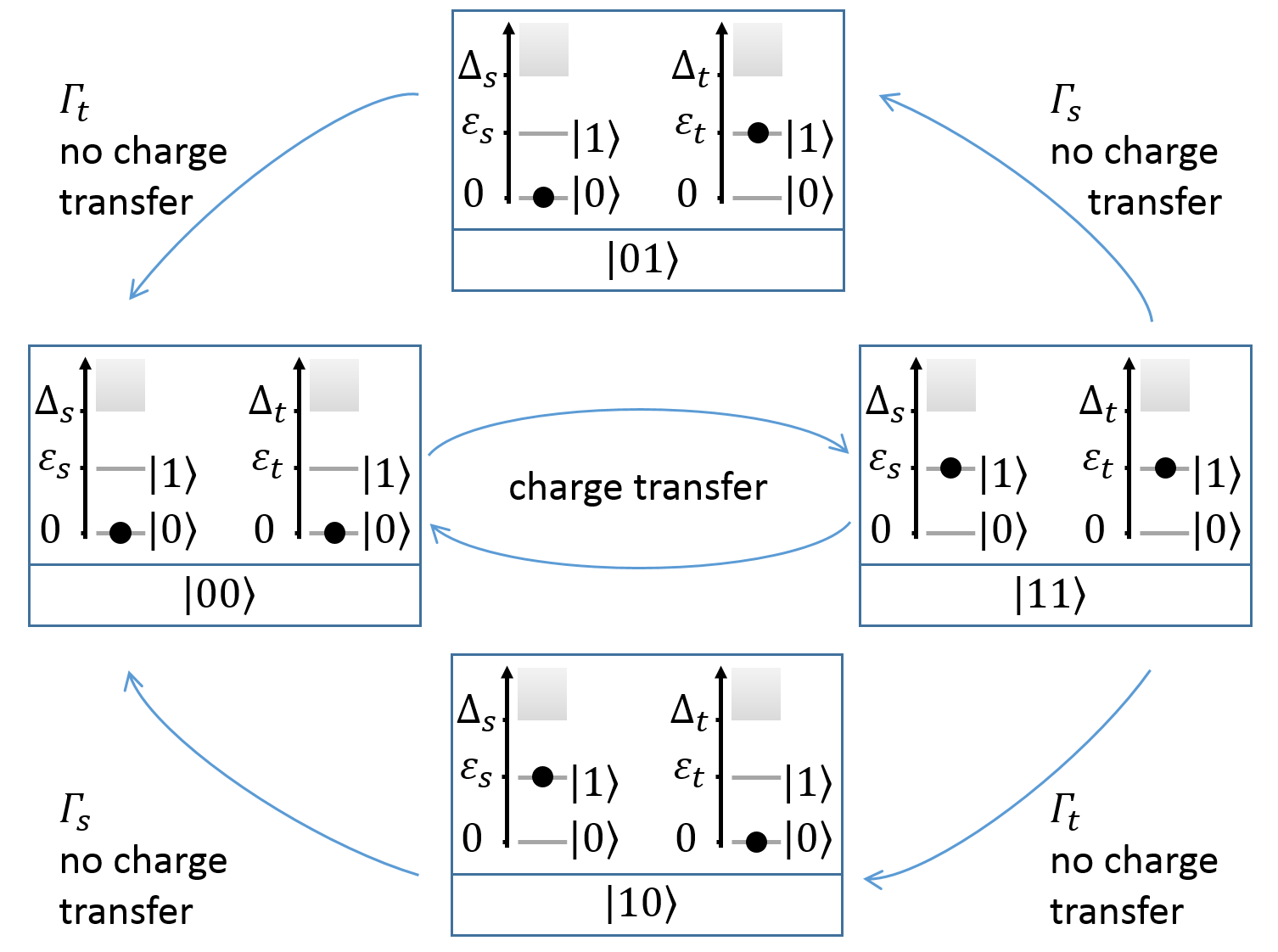}
\caption{Transport diagram. The coherent oscillations of the charge unit $e$ between states $\ket{00}$ and $\ket{11}$
are quenched by incoherent relaxation processes that do not transport charge between the two superconductors. }
\label{fig:sketch5}
\end{center}
\end{figure}

The resulting dynamics is depicted in Fig.~\ref{fig:sketch5}, together with the charge transport associated with each process.

We can easily solve for the steady state, $\dot{\rho}=0$. The resulting steady state probabilities $P_{00}$ and $P_{11}$ are
\begin{align}
P_{11} = & \left(\frac{(\Gamma_\text{s}+\Gamma_\text{t})^2}{\Gamma_\text{s}\Gamma_\text{t}} + \frac{(\Gamma_\text{s}+\Gamma_\text{t})^2}{4 |\gamma_e|^2} + \frac{ (ev)^2}{|\gamma_e|^2}\right)^{-1};\label{eq:P11}\\
P_{01} = &\; (\Gamma_\text{s}/\Gamma_\text{t})P_{11}; \quad P_{10} = (\Gamma_\text{t}/\Gamma_\text{s})P_{11};\notag\\
P_{00} = & \left(1-P_{01}-P_{10}-P_{11}\right).\notag
\end{align}
Their form is particularly simple assuming $\Gamma_\text{s}=\Gamma_\text{t}=\Gamma$,
\begin{align}
P_{00} = & \left(1-P_{01}-P_{10}-P_{11}\right); \notag\\
P_{01} = & P_{10} = P_{11} = \frac{|\gamma_e|^2}{\Gamma^2 + 4 |\gamma_e|^2 + (ev)^2}. \notag
\end{align}

\subsubsection*{Tunneling current}

To find the steady state current, we note that a charge $e$ is transported
across the bias with each coherent excitation $\ket{00}\rightarrow\ket{11}$ that relaxes incoherently.
Therefore, the current is given by
the steady state probability $P_{11}$, Eq.~\ref{eq:P11}, and the total effective lifetime broadening
$\Gamma_\text{eff} = 2\Gamma_\text{s}\Gamma_\text{t}/(\Gamma_\text{s}+\Gamma_\text{t})$,
\begin{align}
I_0(v) = & \frac{e}{\hbar} P_{11} \Gamma_\text{eff}\\
= & \frac{e}{\hbar}\left(\frac{2\Gamma_\text{s}\Gamma_\text{t}}{\Gamma_\text{s}+\Gamma_\text{t}}\right)\frac{{|\gamma_e|}^2}{{|\gamma_e|}^2\frac{(\Gamma_\text{s}+\Gamma_\text{t})^2}{\Gamma_\text{s}\Gamma_\text{t}}+\frac{(\Gamma_\text{s}+\Gamma_\text{t})^2}{4}+(ev)^2},
\label{eq:eq_si_1}
\end{align}
where $ev=eV\pm|\varepsilon_\text{s}+\varepsilon_\text{t}|$ and $|\gamma_e|\propto\sqrt{G_\text{N}}$.
We obtain for $I_0(v)$ a simple Lorentzian function, derived without accounting for environmental broadening.

The effect of the environment broadens the Lorentzian peak, but preserves its area,
\begin{equation}
A=\int^{+\infty}_{-\infty} I_0(v)dv=\frac{4\pi}{\hbar}\frac{(\Gamma_\text{s}\Gamma_\text{t})^{3/2}}{(\Gamma_\text{s}+\Gamma_\text{t})^2}%
\frac{{|\gamma_e|}^2}{\sqrt{{4|\gamma_e|}^2+\Gamma_\text{s}\Gamma_\text{t}}}
\label{eq:eq_si_2}
\end{equation}
For equal lifetime broadening $\Gamma_\text{s}=\Gamma_\text{t}=\Gamma$, we find the simpler expressions,
\begin{align}
I_0(v) = & \frac{e}{\hbar} \Gamma \frac{|\gamma_e|^2}{\Gamma^2 + 4 |\gamma_e|^2 + (ev)^2}.\\
A = & \frac{\pi}{\hbar}\Gamma\frac{{|\gamma_e|}^2}{\sqrt{{4|\gamma_e|}^2+\Gamma^2}}
\end{align}

\subsection*{Environmental Interaction}

The line width of the Shiba-Shiba peak in Fig.\ 4(b) of the main text does not reflect the intrinsic lifetime broadening of the YSR states. We have to take into account the interaction with the environment through a convolution with the $P(E)$ function, which does not change the area of the Shiba-Shiba peak, but certainly its width.

The $P(E)$ function is a probability function describing the probability for an exchange of energy $E$ of the tunneling electron with the environmental impedance during the tunneling process. It can be straightforwardly calculated numerically by assuming an environmental impedance along with a junction capacitance and temperature, which has been described elsewhere \cite{si_Ingold1992,si_Ingold1994,si_Ast2016}. For simplicity, we use an ohmic impedance, which only features one free parameter.

\begin{figure}[ht]
\begin{center}
\includegraphics[width=\columnwidth]{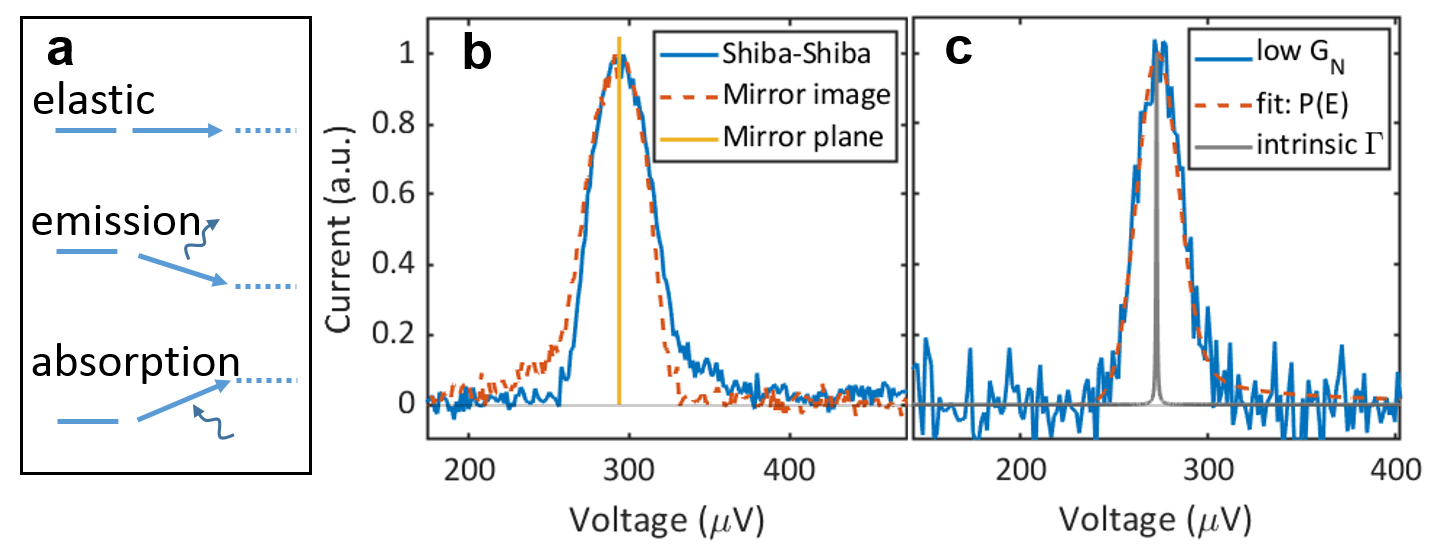}
\caption{Broadening of the Shiba-Shiba peak due to the interaction with the environment during tunneling. \textbf{(a)} Schematic of the tunneling process in resonance and off resonance. \textbf{(b)} Shiba-Shiba peak with its mirror image showing the asymmetry of the $P(E)$ function. \textbf{(c)} Fit of the $P(E)$ function to the Shiba-Shiba peak along with the extremely narrow intrinsic line width for comparison.}
\label{fig:env}
\end{center}
\end{figure}

The interaction of the tunneling electron with the environment during the Shiba-Shiba process is visualized in Fig.\ \ref{fig:env}(a). Detuning the energy levels by changing the bias voltage forces a gain or loss of energy into the environment in order to complete the tunneling process due to the extremely sharp intrinsic line widths of the YSR states. In this sense, the Shiba-Shiba peak actually presents a direct measurement of the $P(E)$ function, which renders Shiba-Shiba tunneling as both quantum sensor and emitter. The characteristic asymmetry of the $P(E)$ function due to detailed balance can be directly seen by superimposing the mirror image as shown in Fig.\ \ref{fig:env}(b). A fit of the Shiba-Shiba peak with the $P(E)$ function is shown in Fig.\ \ref{fig:env}(c) with a FWHM around $27\,\upmu$V. We find very good agreement with the data using a temperature of 11\,mK and a junction capacitance of 5.2\,fF as well as an ohmic impedance of 377\,$\upOmega$.

\subsection*{Lifetime extraction}

In reality, the measured spectrum $I(v)$ has to be convolved with the $P(E)$ function, which contains the environmental broadening \cite{si_Ast2016}:
\begin{equation}
I(v)=\int^{+\infty}_{-\infty} I_0(v)P(e(v-v'))dv'.
\label{eq:eq_si_3}
\end{equation}

Since the convolution conserves the area, the information contained in the area is unbiased by the interaction with the environment.

\textit{Linear regime}. When the tunnel coupling is small, ${|\gamma_e|}^2\ll\frac{\Gamma_\text{s}\Gamma_\text{t}}{4}$, the area is linearly proportional to the conductance
\begin{equation}
A\simeq\frac{4\pi}{\hbar}\frac{\Gamma_\text{s}\Gamma_\text{t}}{(\Gamma_\text{s}+\Gamma_\text{t})^2}{|\gamma_e|}^2\propto G_\text{N}.
\end{equation}

\textit{Sublinear regime}. When the tunnel coupling is large, ${|\gamma_e|}^2\gg\frac{\Gamma_\text{s}\Gamma_\text{t}}{4}$, the area is proportional to square root of conductance, which we call the \textit{sublinear regime}
\begin{equation}
A\simeq\frac{2\pi}{\hbar}\frac{(\Gamma_\text{s}\Gamma_\text{t})^{3/2}}{(\Gamma_\text{s}+\Gamma_\text{t})^2}{|\gamma_e|}\propto \sqrt{G_\text{N}}.
\end{equation}

In the linear regime, the tunnel coupling is the limiting factor, and the intrinsic relaxation is fast enough to relax the excited state so that the next quasiparticle does not see any blockade caused by the previous tunneling event. In the sublinear regime, the tunnel coupling is strong enough that quasiparticles cannot relax fast enough before next tunneling attempt; a coherent oscillation is formed inside the junction. The linear to sublinear transition offers a reliable way to estimate the lifetime, as will be discussed in the following section.

In the case of $\Gamma_\text{s}=\Gamma_\text{t}=\Gamma$, Eq. \ref{eq:eq_si_1} and Eq. \ref{eq:eq_si_2} further simplifies to the one in the main text
\begin{eqnarray}
I_0(v) & = & \frac{e\Gamma}{\hbar}\frac{{|\gamma_e|}^2}{4{|\gamma_e|}^2+\Gamma^2+(ev)^2}\\
A & = & \frac{\pi\Gamma}{\hbar}\frac{{|\gamma_e|}^2}{\sqrt{\Gamma^2+4{|\gamma_e|}^2}}
\end{eqnarray}

Without additional broadening, the Shiba-Shiba peak current at the linear to sublinear transition is a direct measurement of intrinsic lifetime. With $P(E)$ broadening, the maximum current will be lowered significantly and is no longer a good measure any more. However, since broadening mechanisms generally preserve the area, we use the following formula to estimate the lifetime directly from the Shiba-Shiba peak area at the linear to sublinear transition point, where ${|\gamma_e|}^2=\frac{\Gamma_\text{s}\Gamma_\text{t}}{4}$:
\begin{equation}
A_\text{trans}=\frac{\pi}{\sqrt{2}\hbar}\left(\frac{\Gamma_\text{s}\Gamma_\text{t}}{\Gamma_\text{s}+\Gamma_\text{t}}\right)^2
\end{equation}

We note that the area at the transition no longer depends on the coherence factors $u$ and $v$, but only reflects the relaxation rate. In the case of $\Gamma_\text{s}=\Gamma_\text{t}=\Gamma$,
\begin{equation}
A_\text{trans}=\frac{\pi}{4\sqrt{2}\hbar}\Gamma^2
\label{eq:eq_si_4}
\end{equation}
Since we cannot separate the lifetimes of the YSR states in tip and sample just from Shiba-Shiba tunneling, we will use Eq.\ \ref{eq:eq_si_4} and treat $\Gamma$ as an effective combined intrinsic broadening. This is also reasonable because tip and sample YSR state are both subject to similar quasiparticle poisoning characterized by the Dynes parameter. Note that $\Gamma$ is in the unit of energy. To convert it to the time domain, we use the following conversion \cite{si_Nazarov1993}:
\begin{equation}
\tau=\frac{h}{\Gamma}
\end{equation}
Consequently, a typical linear to sublinear transition at $1\times 10^{-18}\,$VA (see Fig.\ 4(a) of the main text) can be converted to a broadening $\Gamma = 0.1\,\upmu$eV or a lifetime of $\tau = 41\,$ns.

\subsection*{Relaxation of YSR states at low temperature}
At higher temperature, there will be phonons with enough energy to excite the quasiparticle
to the empty continuum at the energy of $\Delta_\text{s,t}$, thereby allowing it to escape the junction.
However, this mechanism is suppressed
at low temperature ($T\ll 1\,K$), where phonons are depleted due to the exponential dependence
on temperature. To provide some numbers, for a typical YSR state energy of 145\,$\upmu$eV and
a superconducting gap of about 750\,$\upmu$eV (bulk vanadium) we expect at 15\,mK a Boltzmann
factor of about $10^{-48}$ for the thermal excitation of the unoccupied level and a Boltzmann factor
of about $10^{-203}$ for the thermal relaxation of the excited quasiparticle into the continuum.
Therefore, we can conclude that thermally activated processes can be neglected at low temperature (15\,mK).

We attribute the finite relaxation lifetime of excited YSR states to excess quasiparticles in the gap.
Excess quasiparticles recombine with the excited quasiparticles, thereby changing the quasiparticle parity. Although excess quasiparticles are exponentially suppressed in equilibrium at 15\,mK, a finite density of out of equilibrium quasiparticles has been repeatedly observed in superconducting systems, both in the dynamics of Andreev bound states \cite{si_Janvier2015}
and in superconducting resonators \cite{si_deVisser2011}. The possible sources of out of equilibrium quasiparticles
at low temperatures are still an open question of considerable interest and a subject of debate.

\subsection*{Thermal Shiba-Shiba tunneling}

In this section we describe tunneling between states in the odd-parity manifold.
We proceed analogously to the direct Shiba-Shiba tunneling, parametrizing the
voltage near the thermal Shiba-Shiba resonance by $eV=E_1-E_2+ev$, with the small
detuning $ev \ll E_1,E_2$. The dynamics between odd states
$\ket{01}$ and $\ket{10}$ is described by the Hamiltonian,
\begin{align}
H_o = \begin{bmatrix}
-ev/2 & \gamma_o \\
\gamma_o^* & ev/2
\end{bmatrix},\quad \gamma_o=\sandwich{{10}}{H_\text{T}}{{01}},
\label{Ho}
\end{align}
where $\gamma_o$ is the tunneling matrix element between odd-parity states.

At low temperature, the excited states are unpopulated and the transport resonance is absent.
At finite temperature, we must take into account the rate of thermal excitation of the isolated
YSR states, $\Gamma^e_{i}$. The rate of thermal excitation is reduced compared to the relaxation
rate by the Boltzmann factor $\Gamma^e_{i} = \Gamma_{i} \exp(-k_\text{B}T/2\varepsilon_i)$.

Including the rates of thermal excitation, the dynamics is governed by the following master equation
\begin{align}
\hbar\dot\rho = & i [H_o,\rho] + \\
                 \ & \left[\Gamma^e_\text{s} P_{01}+\Gamma^e_\text{t} P_{10}-(\Gamma_\text{s}+\Gamma_\text{t}) P_{11}\right] \ket{11}\bra{11} + \notag\\
                 \ & \left[(\Gamma_\text{s} P_{11}+\Gamma^e_\text{t} P_{00})-(\Gamma_\text{t} +\Gamma^e_\text{s}) P_{01}\right] \ket{01}\bra{01} + \notag\\
				 \ & \left[(\Gamma_\text{t} P_{11}+\Gamma^e_\text{s} P_{00})-(\Gamma_\text{s}+\Gamma^e_\text{e}) P_{10}\right] \ket{10}\bra{10} +\notag\\
				 \ & \left[(\Gamma_\text{s} P_{10}+\Gamma_\text{t} P_{01})-(\Gamma^e_\text{s}+\Gamma^e_\text{t}) P_{00}\right]\ket{00}\bra{00} -\notag\\
				 \ & \frac{1}{2}(\Gamma_\text{s}+\Gamma_\text{t}+\Gamma^e_\text{s}+\Gamma^e_\text{t})
                 \sum_{x \neq y}\left(\rho_{xy}\ket{x}\bra{y}+\text{h.c.}\right).\notag
\end{align}
We linearize in the small thermal excitation rates $\Gamma^e_{i}$ and find the steady state,
\begin{align}
P_{11} = & 0 + {\mathcal O}(\Gamma^{e\,2}_{i});\\
P_{01} = & \left[\frac{\Gamma_\text{th}^2-(\delta\Gamma_\text{st})^2}{\Gamma_\text{th}^2+(ev)^2} + \frac{(ev)^2}{\Gamma_\text{th}^2+(ev)^2}\right] \frac{\Gamma^e_\text{t}}{\Gamma_\text{t}};\notag\\
P_{10} = & \left[\frac{\Gamma_\text{th}^2-(\delta\Gamma_\text{ts})^2}{\Gamma_\text{th}^2+(ev)^2} + \frac{(ev)^2}{\Gamma_\text{th}^2+(ev)^2}\right]\frac{\Gamma^e_\text{s}}{\Gamma_\text{s}};\notag\\
P_{00} = & \left(1-P_{01}-P_{10}-P_{11}\right),\notag
\end{align}
where we have introduced the notation $\Gamma_\text{th}$ for the width of the thermal Shiba-Shiba peak,
\begin{align}
\Gamma_\text{th}^2 = \frac{\left(\Gamma_\text{s}+\Gamma_\text{t}\right)^2}{4}+\frac{(\Gamma_\text{s}+\Gamma_\text{t})^2}{\Gamma_\text{s}\Gamma_\text{t}}|\gamma_o|^2,
\end{align}
as well as the asymmetric functions with units of frequency,
\begin{align}
(\delta\Gamma_\text{st})^2 = & \frac{\frac{\Gamma_\text{s} }{\Gamma_\text{t} }-\frac{\Gamma^e_\text{s} }{\Gamma^e_\text{t} }}{1+\frac{\Gamma_\text{s} }{\Gamma_\text{t} }}\frac{(\Gamma_\text{s} +\Gamma_\text{t} )^2}{\Gamma_\text{s} \Gamma_\text{t} }|\gamma_o|^2;\\
(\delta\Gamma_\text{ts})^2 = & \frac{\frac{\Gamma_\text{t} }{\Gamma_\text{s} }-\frac{\Gamma^e_\text{t} }{\Gamma^e_\text{s} }}{1+\frac{\Gamma_\text{t} }{\Gamma_\text{s} }}\frac{(\Gamma_\text{s} +\Gamma_\text{t} )^2}{\Gamma_\text{s} \Gamma_\text{t} }|\gamma_o|^2.
\end{align}
Note that $\delta\Gamma_\text{ts}$ is obtained from $\delta\Gamma_\text{st}$
simply by switching the indexes s and t that label the two superconductors.

\begin{figure}[ht]
\begin{center}
\includegraphics[width=\columnwidth]{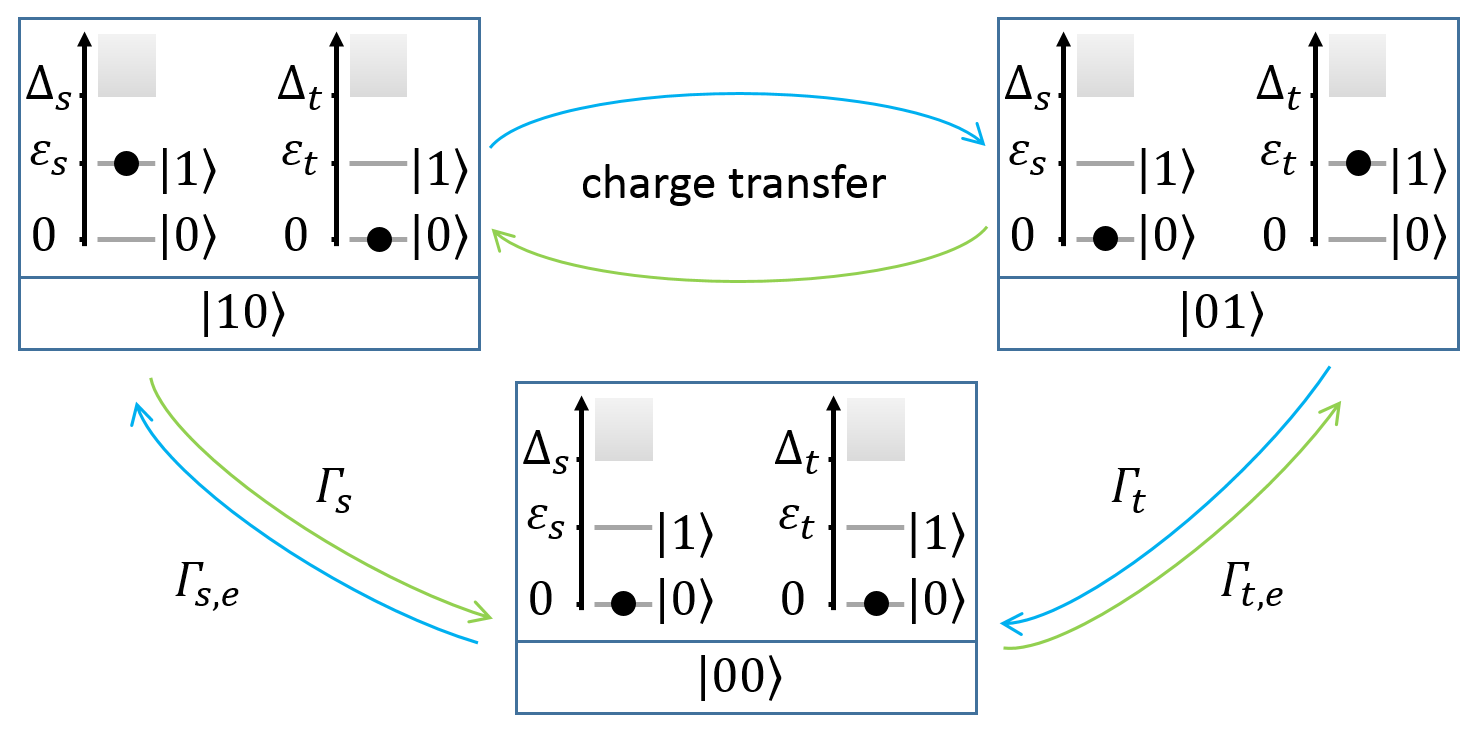}
\caption{(Color online.) Forward and backward transport diagrams showing the elementary
transport processes at the thermal Shiba-Shiba resonance.}
\label{fig:odd}
\end{center}
\end{figure}

The asymmetry is crucial for the thermal Shiba-Shiba resonance, as the current is given by
\begin{align}
I_\text{th} = e\left(P_{01}\Gamma_\text{t}-P_{10}\Gamma_\text{s}\right),
\end{align}
which is a competition between forward and backward electron transfer processes, as depicted in Fig.~\ref{fig:odd}.
For a fully symmetric junction our prediction is that the thermal current will vanish.

The line shape of the thermal Shiba-Shiba resonance is a mixing of Lorentzian and Fano line shapes.

\end{document}